\documentclass[11pt]{article}

\input epsf.sty

\usepackage{graphicx,amsmath,amssymb}
\def\lromn#1{\uppercase\expandafter{\romannumeral#1}}

\begin{document}

\begin{center}
\begin{Large}
\textbf{
Divalent lanthanoid ions  in crystals for neutrino mass spectroscopy
}
\end{Large}

\vspace{1cm}

H. Hara, N. Sasao, and M. Yoshimura

Research Institute for Interdisciplinary Science,
Okayama University \\
Tsushima-naka 3-1-1 Kita-ku Okayama
700-8530 Japan

\vspace{7cm}

{\bf ABSTRACT}

\end{center}

\vspace{1cm}
Electron spin flip in atoms or ions can cause neutrino pair emission, which
provides a method to explore still unknown important neutrino properties by measuring 
spectrum of emitted photon  in association,
when electroweak rates are amplified by a  phase coherence among participating atoms.
Two important remaining neutrino issues to be determined  are the absolute neutrino mass
 (or the smallest neutrino mass in the three-flavor scheme) and the nature of neutrino masses, either
of Dirac type or of Majorana type.
Use of Raman scattered photon was recently proposed as a promising tool
for this purpose.
In the present work we continue along this line to further identify promising ion targets
in crystals, calculate  neutrino
pair emission rates, and study how to extract neutrino properties from Raman scattered photon angular distribution.
Divalent lanthanoid ions  in crystals, in particular Sm$^{2+}$,   are the most promising, due to
(1) its large number density, (2) sharp optical lines, (3) a variety of available ionic levels.
Rejection of amplified quantum electrodynamic backgrounds is made possible
to controllable levels by choosing a range of Raman trigger direction,
when Sm$^{2+}$ sites are at  O$_h$ inversion center of host crystals such as SrF$_2$.

\vspace{5cm}

Keywords
\hspace{0.5cm} 
Neutrino mass,
Majorana fermion,
forced electric dipole transition,
inversion center of crystal point group,
divalent lanthanoid ions in crystals,
Sm$^{2+}$

\newpage

\section
 {\bf Introduction}

Neutrinos are the key particle that can probe physics far beyond the standard theory.
Their finite squared mass differences and mixing in the weak interaction have been discovered 
and determined to
an almost complete level accessible by oscillation experiments \cite{pdg}.
Yet the remaining issues of the absolute mass and the nature of mass term,
either of Majorana or of Dirac type, hold even more important status in the future
of fundamental physics.

We have proposed to use atomic transitions emitting  neutrino pairs along with a photon
in order to determine these remaining important neutrino properties \cite{renp overview}.
The original scheme has more recently been improved 
\cite{ranp} by introducing Raman stimulated process, 
$\gamma_0 + | e \rangle \rightarrow \gamma + | g\rangle + \nu_i \bar{\nu_j}$
($\nu_i\,, i=1,2,3 $ being one of neutrino massive-eigenstate fields),
to distinguish the detected  photon $\gamma$
from otherwise confusing trigger photon $\gamma_0$ by measuring different scattered directions and different
energy.
In the new scheme the scattered  photon angular distribution carries information of 
neutrino properties such as their masses and
Majorana/Dirac distinction.
Both the original and this scheme use a high degree of phase coherence among target atoms.
Amplification of weak process by coherence
has been experimentally verified in quantum electrodynamic (QED) two-photon process
\cite{psr exp}, \cite{psr th}.
The amplification factor was $ \sim 10^{18}$.
The project to determine neutrino properties using atoms or ions with coherence
is called neutrino mass spectroscopy.

In the work \cite{ranp} it was suggested 
to use lanthanoid ions of 4f$^n$ electron system doped  in dielectric crystals.
A great merit of lanthanoid ions as targets is their sharp optical lines at de-excitation
since 4f electrons lying deep inside ions are insensitive to host crystal environment.
The sharpness of optical lines 
can be used to specify resonant intermediate paths in neutrino mass spectroscopy by
high quality laser irradiation.
Lanthanoid ions thus make a compelling case towards successful  Raman stimulated neutrino mass spectroscopy.
The problem of background rejection has been left  unresolved in the work, however.

We  continue in the present paper to work out basics towards  
Raman stimulated neutrino mass spectroscopy, and,
in particular, determine  which lanthanoid ions are most appropriate from the point of background rejection.
The most important condition turns out to be rejection of amplified QED background events.
We find that a point group symmetry endowed with inversion center (in particle physics terminology, parity
conservation holding at the ion site) 
of host crystals greatly helps to reduce QED backgrounds.
The best candidate we found is divalent ion Sm$^{2+}$ at inversion center of O$_h$ symmetric crystals,  doped in
alkali-earth halides such as SrF$_2$ and CaF$_2$ 
(both are transparent crystals in the optical region).
We discuss these cases in detail, and identify the largest amplified QED background, which
turns out well  controllable
by identifying and isolating  emitted extra photons.

We assume for simplicity that  designed experiments are conducted at sufficiently low temperatures,
and ignore finite temperature effects.
The terminology based on the angular momentum conservation  and parity notion in the free space
such as electric dipole and magnetic dipole is used for electron transition operator.
On the other hand, stationary electronic states of ions in crystals are classified in terms of irreducible representation
of crystal point group (which exactly holds), 
but sometimes in terms of approximate Russell-Saunders (or $L-S$) scheme
in atomic physics \cite{atomic physics text}.

The paper is organized as follows.
Section 2 starts from a theoretical formulation applicable to rate calculations
both of macro-coherently amplified neutrino pair emission and QED backgrounds.
Section 3 is devoted to  calculation of rate and angular distribution of  neutrino pair emission,
and Section 4 to QED background events.
In Section 5 we show how angular distributions using Sm$^{2+}$ ion exhibit
important neutrino mass parameters and Majorana/Dirac distinction.
Finally Section 6 presents summary of  the present work and prospects in the future.

We use the natural unit of $\hbar = c = 1$ throughout the present paper unless otherwise stated.
Useful numbers to remember are 1 eV$ = 1.5 \times 10^{15}$sec$^{-1}$ and its inverse $= 1,240$ nm of laser wavelength ,
Avogadro number cm$^{-3} = 7.6 \times 10^{-15}  $eV$^{3}$, $G_F^2\,$eV$^5 = 2.1 \times 10^{-31} $sec$^{-1}$.
Atomic physics uses a unit of energy, cm$^{-1}$, and it is related to eV by $10^{4}$cm$^{-1} = 1.24$eV.

\vspace{0.5cm}
\section %\lromn1
 {\bf Raman stimulated neutrino pair emission: A formulation}

\begin{figure*}[htbp]
 \begin{center}
 \epsfxsize=0.5\textwidth
 \centerline{\includegraphics[width=17cm,keepaspectratio]{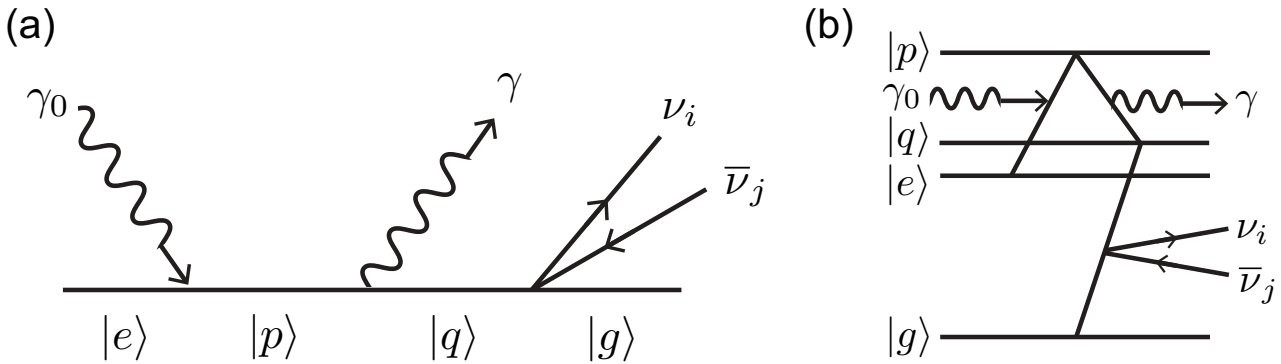}} \hspace*{\fill}
   \caption{
(a) Feynman diagram of  $ \gamma_0 + |e \rangle \rightarrow \gamma +  | g\rangle + \nu_i \bar{\nu_j} $.
There are five more diagrams by changing how three vertexes are arranged, but
at the resonance this diagram is dominant.
 There are five more diagrams that contribute off resonances.
 (b) Corresponding energy levels indicating  absorption and emission of photons and a neutrino-pair.
}
   \label {drenp feynman}
 \end{center} 
\end{figure*}

Suppose that a collective body of atoms/ions de-excite
after Raman scattering as depicted in Fig(\ref{drenp feynman}),  emitting plural particles
which can be either photons or neutrino-pair.
Quantum mechanical transition amplitude and its square of the process, if the phase of atomic part of amplitudes,
 $A_a =A$, is common and uniform, are given by formulas,
\begin{eqnarray}
&&
\sum_a e^{i (  \vec{p}_{eg} + \vec{k}_0 - \vec{k} - \vec{p}_1 - \vec{p}_2 )\cdot \vec{x}_a } A_a \simeq 
 n\, (2\pi)^3 \delta ( \vec{p}_{eg} + \vec{k}_0 - \vec{k} - \vec{p}_1 - \vec{p}_2)\,A
\,,
\label {mc amplitude}
\\ &&
| \sum_a e^{i (  \vec{p}_{eg} + \vec{k}_0 - \vec{k} - \vec{p}_1 - \vec{p}_2 )\cdot \vec{x}_a } 
A(\vec{k}_0,  \vec{k},  \vec{p}_1,  \vec{p}_2 )|^2  \simeq 
n^2V  (2\pi)^3 \delta ( \vec{p}_{eg} + \vec{k}_0 - \vec{k} - \vec{p}_1 - \vec{p}_2)
 | A(\vec{k}_0,  \vec{k},  \vec{p}_1,  \vec{p}_2 )|^2
\,,
\end{eqnarray}
with $n$ the assumed uniform density of excited atoms/ions.
We assumed that atoms/ions are infinitely heavy with no recoil,
hence one does not expect the momentum conservation in the usual stochastic atomic de-excitation.
When a spatial phase coherence exists as in this case, the situation drastically changes, resulting in  the momentum
conservation and rate dependence $\propto n^2 $ of the target number density.
The phase $\vec{p}_{eg} $ is the one imprinted at excitation of atoms/ions by a high quality of lasers.
Equality to the right hand side is valid in the continuous limit of atomic distribution.
The coherence gives rise to a mechanism of  amplification,   realization of two results, 
(1) rate  $\propto n^2 V$ with $V$ the volume of target region, and (2) the momentum conservation.
We call this the macro-coherent (MC) amplification.
Thus, in the macro-coherent Raman stimulated neutrino-pair emission,
 both the energy (as usual) and the momentum conservation
(equivalent to the spatial phase matching condition) hold \cite{renp overview};
\begin{eqnarray}
&&
\omega_0 + \epsilon_{eg} = \omega + E_1 + E_2
\,, \hspace{0.5cm}
\vec{k}_0 + \vec{p}_{eg} = \vec{k} + \vec{p}_1 + \vec{p}_2
\,,
\end{eqnarray}
where $E_i = \sqrt{p_i^2 + m_i^2}$ with  $m_i\,, i =1,2$ of three neutrino masses.
From the energy and the momentum conservation, one may derive the kinetic region
of $(12)$ neutrino-pair emission:
$ ( \omega_0 + \epsilon_{eg} - \omega )^ 2 -  ( \vec{k}_0 + \vec{p}_{eg} - \vec{k} )^ 2 \geq (m_1 +m_2)^2$.
This may be regarded as a restriction to emitted photon energy $\omega$ and its emission angle.
At the location where the equality holds, the neutrino-pair is emitted at rest.
On the other hand, when atomic phases of ${\cal A}_a$ at sites $a$ are random in a given target volume $V$,
the rate scales with $nV$ without the momentum conservation law, which gives
much smaller  rates.

\begin{figure*}[htbp]
 \begin{center}
 \epsfxsize=0.5\textwidth
 \centerline{\includegraphics[width=18cm,keepaspectratio]{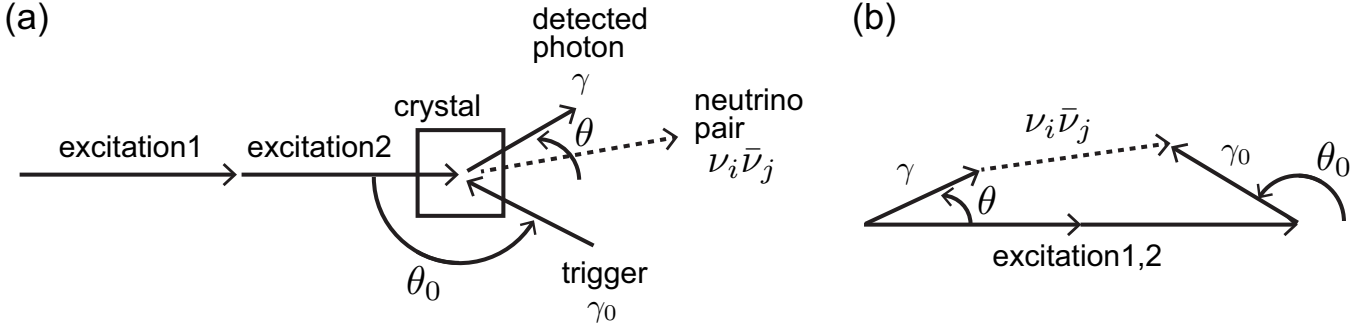}} \hspace*{\fill}
   \caption{
(a) A layout of experimental configuration \hspace{0.3cm}
(b) Momentum conservation due to macro-coherence.
}
   \label {exp layout}
 \end{center} 
\end{figure*}

There are a variety of ways to develop the macro-coherence.
Raman stimulation here was introduced to reduce backgrounds by
taking scattered photon $\gamma$ direction distinct from the trigger photon $\gamma_0$ direction,
as illustrated in an experimental layout in Fig(\ref{exp layout}).

Experimental verification of the principle of macro-coherence was achieved in  QED two-photon emission 
at the vibrational transition of hydrogen molecule \cite{psr exp}, following suggestion of \cite{psr th}.
The achieved enhancement $N_{{\rm eff}}$ reached $\sim 10^{18}$.
We assume below that similar rate amplification works in neutrino process.

As in \cite{ranp}, we assume the resonance condition:
$\omega_0 \approx \epsilon_{pe}\,, \omega \approx \epsilon_{pq} $.
In order to formulate the problem of rate calculation,
we take a two-step picture:  fast Raman scattering of amplitude $\langle q| H_R|e \rangle $ is followed
 by slow weak processes of many
particle emission $\langle g| H_W|q \rangle $  via a long lived state $|q \rangle $.
We derive the transition amplitude at finite times using perturbation theory,
\begin{eqnarray}
&&
{\cal A} =
\int_0^t dt' \int_0^{t\,'} dt'' \langle g| H_W(t') |q \rangle \langle q| H_R( t'')|e \rangle
\,.
\label {raman stimulated weak amplitude}
\end{eqnarray}
Time dependence of  hamiltonian matrix elements is given by
\begin{eqnarray}
&&
\langle g| H_W(t') |q \rangle = e^{-i (- \epsilon_{qg} + E_1+ E_2)t' - \gamma_2 t'/2} H^W_{gq}(0)
\,, \hspace{0.5cm}
\\ &&
\langle q| H_R(t'') |e \rangle = e^{-i (- \epsilon_{eq} + \omega - \omega_0)t'' - \gamma_1 t''/2} H^R_{qe}(0)
\,.
\end{eqnarray}
Time integration of eq.(\ref{raman stimulated weak amplitude}) gives
\begin{eqnarray}
&&
\hspace*{-1cm}
{\cal A} =\frac{H^R_{qe}(0) H^W_{gq}(0)}{ - \epsilon_{eq} + \omega  - \omega_0 - i \gamma_1/2}\left(
\frac{1 - e^{- i (- \epsilon_{eg} + E_1 + E_2 + \omega  - \omega_0 )t -(\gamma_1 + \gamma_2)t/2 }}
{ - \epsilon_{eg} + E_1 + E_2 + \omega - \omega_0 - i (\gamma_1 + \gamma_2)/2} 
+ \frac{1 - e^{- i (\epsilon_{eq} + \omega_0   - \omega )t - \gamma_2 t/2 }}
{  \epsilon_{eq} + \omega_0   - \omega - i  \gamma_2/2} 
\right)
\,.
\nonumber \\ && 
\end{eqnarray}
The transition rate, conveniently defined by the probability per unit time, is hence in general
time dependent function, and is given by
\begin{eqnarray}
&&
\lim_{t \rightarrow \infty} \frac{| {\cal A}  |^2} {t}
=\frac{ | H^R_{qe}(0) H^W_{gq}(0) | ^2}{(- \epsilon_{eq} + \omega - \omega_0 )^2 +\gamma_1^2 /4 } 
\lim_{t \rightarrow \infty} \frac{|B(t)|^2} {t} 
\,,
\\ &&
\hspace*{-1cm}
|B(t)|^2 = 
\left(
\frac{1 +e^{- \gamma_2 t }  -2 e^{- \gamma_2 t/2 } 
\cos (\epsilon_{eq} + \omega_0   - \omega ) t  }
{(\epsilon_{eq} + \omega_0   - \omega )^2 + \gamma_2^2/4 }
\right.
\nonumber \\ &&
\left.
+ \frac{1 +e^{- (\gamma_1+ \gamma_2) t } 
 -2 e^{-  (\gamma_1+ \gamma_2) t/2 } \cos (- \epsilon_{eg} + E_1 + E_2 +  \omega  - \omega_0) t  }
{( - \epsilon_{eg} + E_1 + E_2 +  \omega - \omega_0)^2 + (\gamma_1+ \gamma_2)^2/4 } +
({\rm crossed\; interference\; term})\,
\right)
\,.
\end{eqnarray}

The squared energy denominator in the front, when the Raman trigger $\omega_0$ is fixed at 
$\epsilon_{pe} $, is approximated by
\begin{eqnarray}
&&
\frac{1}{ (- \epsilon_{eq} + \omega - \omega_0   )^2 + (\gamma_e^2 + \gamma_q^2 )/4}
\simeq \frac{ 2\pi \delta (  \omega - \epsilon_{pq} ) }
{\sqrt{ \gamma_e^2 + \gamma_q^2} }
\,,
\end{eqnarray}
with $\gamma_1 = \sqrt{ \gamma_e^2 + \gamma_q^2}$,
This gives a finite time behavior of transition probability,
\begin{eqnarray}
&&
\frac{|{\cal A} (t)|^2} {t} \simeq
\frac{2\pi \delta (- \epsilon_{eq} + \omega - \omega_0 )  | H^R_{qe}(0) H^W_{gq}(0) | ^2}
{ \gamma_1 \, t} 
\left(
\frac{1 +e^{- \gamma_2 t }  -2 e^{- \gamma_2 t/2 } 
\cos (\epsilon_{eq} + \omega_0   - \omega ) t  }
{( \epsilon_{eq} + \omega_0   - \omega )^2 + \gamma_2^2/4 }
\right.
\nonumber \\ &&
\left.
+ \frac{1 +e^{- (\gamma_1+ \gamma_2) t } 
 -2 e^{-  (\gamma_1+ \gamma_2) t/2 } \cos (- \epsilon_{eg} + E_1 + E_2 +  \omega  - \omega_0) t  }
{( - \epsilon_{eg} + E_1 + E_2 +  \omega - \omega_0)^2 + (\gamma_1+ \gamma_2)^2/4 }
+
({\rm crossed\; interference\; term})
\right)
\,.
\end{eqnarray}
The first and interference terms in the right hand side of this equation give  irrelevant processes to
neutrino-pair emission, hence are disregarded hereafter.

\begin{figure*}[htbp]
 \begin{center}
 \epsfxsize=0.6\textwidth
 \centerline{\includegraphics[width=10cm,keepaspectratio]{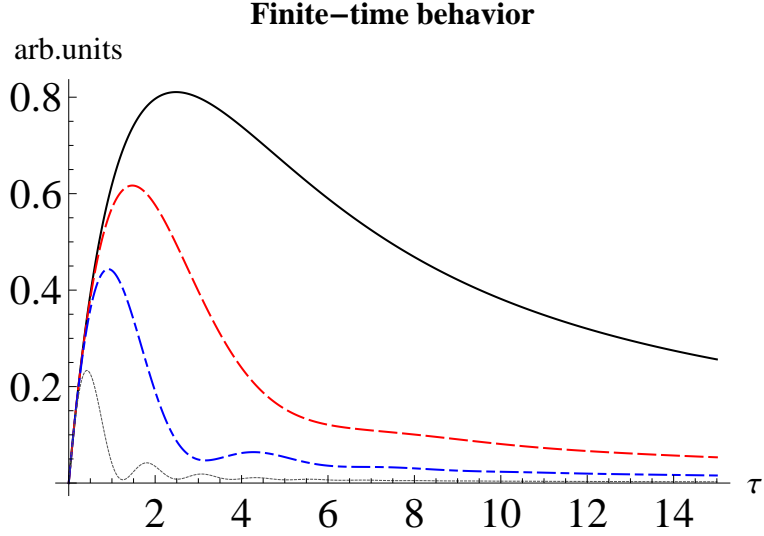}} \hspace*{\fill}
   \caption{ 
Finite time behavior of transition probability per unit time.
$| E_1 + E_2 - \epsilon_{qg}|/ $decay width $ =$ 0.1 in solid black,
1 in dashed red, 2 in dash-dotted blue, and 5 in dotted black.
The most detuned case shown by the dotted black curve gives the
smallest stationary rate, while
the initial  (at $t \times$  width $< O(5)$) oscillating behavior is most conspicuous.
}
   \label {t-evolution}
 \end{center} 
\end{figure*}

The finite time behavior derived here according to quantum mechanical rules is  given in terms
of dimensionless time $\tau$ and energy scaled by decay rate $\gamma$,
\begin{eqnarray}
&&
\frac{1 + e^{- \tau} - 2   e^{- \tau/2} \cos (\Delta \tau )} {\tau \left(\Delta^2 + 1/4 \right) }
\,, \hspace{0.5cm}
\tau = \gamma t
\,, \hspace{0.5cm}
\Delta = \frac{ (- \epsilon_{eg} + E_1 + E_2 +  \omega - \omega_0 ) }{\gamma }
\,,
\end{eqnarray}
with $\gamma = \gamma_1 + \gamma_2 $.
This function is illustrated for several values of $\Delta$ in Fig(\ref{t-evolution}).
The dimensionless quantity $\Delta$ is usually taken much larger than unity, $\Delta \gg 1$, giving the Fermi golden rule,
\begin{eqnarray}
&&
\lim_{ \tau \rightarrow \infty} \frac{ 4 \sin^2 \frac{ \Delta \tau}{2 } } {\tau \Delta^2 } \simeq 2\pi \delta(\Delta)
\,,
\end{eqnarray}
equivalent to the stationary probability per unit time.
A typical lifetime in the problem we discuss in the present work is  of order 1 msec, much larger
than frequency period of infrared (IR) light wave,
and we shall assume the stationary decay rates hereafter.
Nevertheless, it would be interesting to observe finite oscillatory behavior if possible.

By taking a time region of $1/ | - \epsilon_{eg} + E_1 + E_2 +  \omega  - \omega_0 |\ll  t \ll 1/(\gamma_1 + \gamma_2) $, 
one is led to the formula according to the Fermi rule,
\begin{eqnarray}
&&
\frac{|{\cal A} (t)|^2} {t} \simeq
\frac{ (2\pi)^2 \delta (- \epsilon_{eq} + \omega - \omega_0 )  | H^R_{qe}(0) H^W_{gq}(0) | ^2}
{ \gamma_1 } 
 \delta(- \epsilon_{eg}  + E_1 + E_2 + \omega -\omega_0 )
\,.
\end{eqnarray}
We may identify $2\pi   | H^R_{qe}(0)  |^2\,\delta (- \epsilon_{eq} + \omega - \omega_0 ) $ as Raman rate, 
$2\pi   | H^W_{gq}(0) |^2\,\delta(  \epsilon_{eg} + \omega_0  - E_1 - E_2 - \omega ) $ as weak rate.
The factor $1/\gamma_1  $ is the lifetime of intermediate state.

Hence, rate of Raman stimulated weak process is understood as a product of three factors,

(MC amplified) Raman rate $\times$ lifetime of intermediates state $\times$ weak rate.\,

or is given by the formula,
\begin{eqnarray}
&&
\frac{|{\cal A} (t)|^2} {t} \simeq d\Gamma_R \frac{1}{  \sqrt{ \gamma_e^2 + \gamma_q^2}} d\Gamma_W
\,, 
\\ &&
d\Gamma_R = 2\pi  | H^R_{qe}(0)  |^2\,\delta ( \epsilon_{eq} + \omega_0 - \omega )
\,, \hspace{0.5cm}
d\Gamma_W =  2\pi | H^W_{gq}(0) |^2\,\delta(  \epsilon_{qg}  - E_1 - E_2  )
\,.
\end{eqnarray}
This product decomposition assures that Raman and weak processes can be separately discussed,
and we may simply multiply two rates and the lifetime factor in the end.
This makes it possible to calculate rates of neutrino-pair emission and QED backgrounds on the same footing.

We first discuss the Raman rate $d\Gamma_R $.
Convolution of the narrow resonance function above 
with a Lorentzian Raman trigger power spectrum $f(\omega_0)E_0^2$ gives
\begin{eqnarray}
&&
\int d\omega_0 \frac{ f(\omega_0) E_0^2} 
{ (\omega_0 -W)^2 + (\gamma_e^2 + \gamma_q^2)/4 }G(\omega_0)
\approx  \frac{4 E_0^2 }{ \Delta \omega_0  \sqrt{ \gamma_e^2 + \gamma_q^2 } }G(W) 
\,,
\end{eqnarray}
(with $\Delta \omega_0 \gg  \sqrt{ \gamma_e^2 + \gamma_q^2 }$)
provided that the resonance point at $\omega_0 =W $ is within the support range of the spectrum function.
Here $G(\omega_0)$ is a function varying more slowly than the spectrum function $f(\omega_0)$.
This relation is valid for a Lorentzian shape $f(\omega_0)$
and for a Gaussian shape the pre-factor is slightly changed: $4 \rightarrow 2\sqrt{\pi} $.
A numerical difference amounts to  1.13, which is unimportant to accuracy of our following results.
We shall use the Lorentzian result hereafter.

The momentum conservation holds for the whole process, but one can conveniently
insert an extra momentum integration as
\begin{eqnarray}
&&
\delta (\vec{p}_{eg} + \vec{k}_0 - \vec{k} - \vec{p}_1 - \vec{p}_2 )
= \int d^3 P\, \delta (\vec{p}_{eg} + \vec{k}_0 - \vec{k} - \vec{P} ) 
\delta (\vec{P} - \vec{p}_1 - \vec{p}_2 )
\,.
\end{eqnarray}
Although the $n^2 V$ MC amplification factor holds for the whole process once,
one may assume the momentum conservation for each sub-process at will.
The Raman scattering amplitude is 
\begin{eqnarray}
&&
{\cal A}_R = \frac{\vec{d}_{pe}\cdot \vec{E}_0 \vec{d}_{pq}\cdot \vec{E} }{\epsilon_{pe}- \omega_0  - i (\gamma_p + \gamma_e)/2}
\,.
\end{eqnarray}
The squared electric dipole (E1) or magnetic dipole (M1) transition moments,
$d_{ab}^2$ or $\mu_{ab}^2$, is both expressed in terms of
rate divided by level spacing to the third power, $3 \pi \gamma_{ab}/\epsilon_{ab}^3$.
The Raman scattering  rate, including  the MC amplification, is then given by
\begin{eqnarray}
&&
d\Gamma_R =  \frac{72 \pi^3 \gamma_{pe } \gamma_{pq }  d\Omega}{ \epsilon_{pe }^3  \epsilon_{pq }^2} 
\frac{n^2 V E_0^2 | \rho_{eq}|^2  
}{ \Delta \omega_0 \, ( \gamma_p^2 + \gamma_e^2) } 
\,.
\end{eqnarray}
The parameter $| \rho_{eq}|^2 $ is  related to
a generated coherence between the  initial and final states, 
and in general time dependent: $| \rho_{eq} (t) |^2 $.
In a purely quantum mechanical system without dissipation $\rho_{eq} = c_e^* c_q$
with $c_a$ the probability amplitude in state $|a \rangle$,
but with dissipation this definition is extended to the density matrix element in dissipative system that
follow complicated equations.
Systematic calculation of time varying $\rho_{eq} (t) $ is beyond the scope of the present work,
but sample calculations of $\rho_{eq} (t) $ based
on the Maxwell-Bloch equation, a coupled system of non-linear and partial differential equations
for fields and density matrix elements,  
are given in \cite{psr th}, \cite{renp overview}.
We introduce a factor $\eta$ defined by
$ E_0^2 |\rho_{eq }|^2= \eta \epsilon_{pe} n/2 $ for convenience.

Typical values valid in the present work 
of level spacing, $\epsilon_{pe} = \epsilon_{pq }= 100 \sim 500 \,$meV,
lifetimes $\gamma_{pe } =\gamma_{pq } =1/$msec,  and 
$\Delta \nu_0 = \Delta \omega_0/2\pi  = 1 {\rm GHz}  \gg \gamma_i$
give an event number (rate $\times$ duration time $\Delta t$ taken as the lifetime), 
\begin{eqnarray}
&&
\hspace*{-1cm}
\frac{ d\Gamma_R \Delta t}{d\Omega}
 = \left( 4.8  \times10^{46} \sim 7.6 \times 10^{43} \right) \, 
\frac{ 1 {\rm GHz}}{ \Delta \nu_0 }\, \eta
(\frac{n }{  10^{15} {\rm cm}^{-3}})^3\frac{V}{ 10^{-2} {\rm cm}^3}
\frac{ (10^3{\rm sec}^{-1})^3 }{ ( \gamma_p^2 + \gamma_e^2)  \sqrt{\gamma_e^2 + \gamma_q^2 }}
\,.
\label {mc raman rate}
\end{eqnarray}
The choice of $n^3 V = (10^{15})^3 \times 10^{-2} {\rm cm}^{-6} $ is understood as follows.
Consider  irradiated length of crystal target,  1 cm, with
1 \% dopant ions corresponding to target density of order $10^{20}$cm$^{-3}$,
hence a total target number $10^{18}$ for $V=10^{-2}$cm$^3$.
1 mJ laser can in principle excite a number $10^{16}/1.6$ atoms, hence CW (Continuous Wave) operation
of  1 mm$^2$  focused region creates $10^{18}/1.6$  excited ions per unit volume with 100 \% efficiency.
We took three orders reduction factor of $n  $ with the 1  mm$^2$ focusing effect included.
This may or may not be an ideal excitation, 
and the efficiency given by $ | \rho_{eq}|^2$ may give a further reduction.

This large enhancement, eq.(\ref{mc raman rate}), 
encourages us to examine Raman stimulated neutrino pair emission
towards atomic spectroscopy of neutrino masses.

Raman scattering process creates atomic states of well-defined energy 
$E = \epsilon_{eg} + \omega_0 - \omega $ and spatial phase
or momentum $\vec{P}= \vec{p}_{eg} + \vec{k}_0 - \vec{k} $ common to all
weak processes, including higher order amplified QED processes.
The atomic system behaves as if it has an energy-momentum $(E,  \vec{P})$.
Coherently excited $|q \rangle $ states by Raman process
 then de-excite and they may emit weakly interacting light particles
such as neutrino-pairs.
MC amplified Raman stimulated rates of all these processes are given by 
\begin{eqnarray}
&&
d\Gamma_{RW} =  4 \frac{d\Gamma_R d\Gamma_W }{ \sqrt{\gamma_e^2 + \gamma_q^2} }
\,.
\end{eqnarray}
This is a master formula used for subsequent RANP rate and background rate calculations.

\vspace{0.5cm}
\section
{\bf Neutrino pair emission rate}

We next turn to the neutrino pair emission rate, the  case of $\Gamma_W =  \Gamma_{2\nu}$.
Neutrino pair is emitted by electron spin flip given by the electron spin operator $\vec{S}$.
Its coupling in hamiltonian with the neutrino pair current
$ \vec{{\cal N}_{ij}} = \nu_i^{\dagger} \vec{\sigma}\nu_j $ (in the two-component formalism)
produces a neutrino pair  $ (ij)$ of mass eigenstates.
Total neutrino pair emission rate $\Gamma_{2\nu} $  
has been calculated  \cite{ranp} in terms of
squared mass  function 
${\cal M}^2( \theta) = (\epsilon_{eg} + \omega_0 - \omega )^2 - (\vec{p}_{eg} + \vec{k}_0 - \vec{k} )^2$:
the necessary definitions and results are
\begin{eqnarray}
&&
{\cal H}_W = \frac{G_F}{\sqrt{2}} \sum_{ij} b_{ij}\, e^{\dagger} \vec{\sigma} e \cdot \vec{{\cal N}_{ij}}
\,, \hspace{0.5cm}
b_{ij} = U_{ei}^*U_{ej} - \frac{1}{2} \delta_{ij}
\\ &&
\Gamma_{2\nu} =\frac{ G_F^2}{2} n_0 \sum_{ij}  F_{ij} ( \theta)
\Theta \left( {\cal M}^2 ( \theta) - (m_i +m_j )^2\right) 
\,, 
\label {diff rate0}
\\ &&
{\cal M}^2  ( \theta)
=  2 \epsilon_{eg} \left( \epsilon_{pe} (1 - \cos \theta_0)  - \epsilon_{pq} ( 1- \cos \theta) 
\right) - 4 \epsilon_{pe}  \epsilon_{pq} \sin^2 \frac{\theta - \theta_0}{2}
\,,
\label {squared mass f}
\\ &&
 F_{ij} (\theta)
=  \int \frac{d^3p_1 d^3p_2} { (2\pi)^2} \delta (\epsilon_{eg} + \epsilon_{pe} - \epsilon_{pq} - E_1-E_2)  
\delta (\vec{p}_{eg} - \vec{k} - \vec{p}_1 - \vec{p}_2)
\vec{{\cal N}_{ij}}\cdot \vec{ {\cal N}_{ij}}^{\dagger}
\,.
\label {diff rate}
\end{eqnarray}
The $3 \times 3$ unitary matrix $(U_{ei} )\,, i = 1,2,3,$ refers 
to the neutrino mass mixing \cite{pdg}.
Angles, $\theta_0 \,, \theta$, are measured from the excitation axis parallel to $\vec{p}_{eg}$.
From the macro-coherence condition the squared mass function
is equal to the invariant squared mass of the neutrino-pair system:
${\cal M}^2( \theta) = (p_1+p_2)^2 $.

The factor $n_0$ in eq.(\ref{diff rate0}) is derived as follows.
The macro-coherence amplification factor $n^2 V $ applies to the total
RANP process, but for convenience we included it in the Raman rate $\Gamma_R$.
To respect the momentum conservation in the following step $|q \rangle \rightarrow | g \rangle$,
we manipulate the formula to insert an identity equivalent to eq.(\ref{mc amplitude}) with ${\cal A}=1$,

\begin{eqnarray}
&&
n (2\pi)^3 \delta ( \vec{p}_{eg} + \vec{k}_0 - \vec{k} - \vec{p}_1 - \vec{p}_2 = 0 )
= \int_{V} \frac{d^3 x}{V}\, e^{i \vec{0}\cdot \vec{x}} =1
\,,
\end{eqnarray}
from which we derive $ (2\pi)^3 \delta ( \vec{p}_{eg} + \vec{k}_0 - \vec{k} - \vec{p}_1 - \vec{p}_2 ) = 1/n$, hence
$n_0 = n$.

The step function 
 $\Theta\left( {\cal M}^2 (\theta)
- (m_i +m_j )^2\right)$ determines locations of  $(ij)$ neutrino-pair production thresholds.
The squared neutrino pair current $\vec{{\cal N}_{ij}}\cdot \vec{ {\cal N}_{ij}}^{\dagger} $ 
is summed over neutrino helicities and their momenta:
\begin{eqnarray}
&&
\sum_{{\rm helicities}} \vec{{\cal N}}_{ij} ^{\dagger}\cdot\vec{{\cal N}}_{ij} = 
\frac{1}{2} \left( 1 - \frac{ \vec{p}_1 \cdot \vec{p}_2}{E_1 E_2} - \delta_M \frac{m_i m_j }{E_1 E_2 }
\right)
\,,
\label {helicity sum}
\\ &&
\vec{p}_1 \cdot \vec{p}_2 = \frac{1}{2} \left( ( \vec{p}_1 + \vec{p}_2)^2 - \vec{p}_1^2 - \vec{p}_2^2 \right)
=  \frac{1}{2} (\epsilon_{qg}^2 -   {\cal M}^2) - \frac{1}{2} (E_1^2 + E_2^2 - m_i^2 - m_j^2 )
\nonumber \\ &&
= E_1 E_2 - \frac{1}{2} ({\cal M}^2 - m_i^2 - m_j^2)
\,.
\end{eqnarray}
The important formula, eq.(\ref{helicity sum}), was derived for the first time in \cite{my-07}.
The quantity $ F_{ij} $ that appears in the formula, eq.(\ref{diff rate}), is calculated as %\cite{mistake}
\begin{eqnarray}
&&
F_{ij} ( \theta)
= \frac{1}{8\pi} 
\left\{
\left(1 - \frac{ (m_i + m_j)^2}{  {\cal M}^2( \theta)} \right)
\left(1 - \frac{ (m_i - m_j)^2}{  {\cal M}^2 ( \theta)} \right)
\right\}^{1/2} 
\nonumber \\ &&
\times
\left[ \frac{1}{2} |b_{ij}|^2
\left({\cal M}^2 ( \theta) - m_i^2 - m_j^2  \right)
- \delta_M \, \Re( b_{ij}^2 ) \,m_i m_j \right]
\,.
\label {2nu integral}
\end{eqnarray}
$\delta_M =0 $ for Dirac neutrino and $=1 $ for Majorana neutrino due to identical fermion effect \cite{my-07}.

Let us assume, in order to  quantify the Majorana/Dirac distinction,
 that $b_{ii} = 1/2$ for a single $i$ neutrino-pair emission in the rate formula, 
and work out difference of rates in
Dirac pair and Majorana pair emission.
At $(ii)$ threshold ${\cal M}^2 ( \theta) = 4 m_i^2 $ and the difference near threshold is
\begin{eqnarray*}
&&
16\pi F_{ii}   ( \theta)
= \left(1 - \frac{ 4 m_i^2}{  {\cal M}^2( \theta)} \right)^{1/2} \frac{1}{4}
\left( {\cal M}^2 ( \theta) - 4m_i^2 \right)
\sim \frac{ p^3}{m_i} \; ({\rm Majorana \; pair})
\,, 
\\ &&
16\pi F_{ii}  ( \theta)
=  \left(1 - \frac{ 4 m_i^2}{  {\cal M}^2( \theta)} \right)^{1/2} \frac{1}{4}
\left( {\cal M}^2 ( \theta) - 2m_i^2  \right)
 \sim \frac{1}{2} m_i p \; ({\rm Dirac\; pair})
\,,
\end{eqnarray*}
with ${\cal M}^2 ( \theta) = 4(p^2+m_i^2) $ and $p$ the relative momentum of neutrino-pair.
This difference means that the Majorana-pair emission has
an extra suppression factor of momentum $p^2/m_i^2$ at the threshold. 
It should be clear that this is caused by the  P-wave nature of identical fermion
pair emission in the Majorana case giving rise to $\propto p^3$, while the Dirac-pair emission occurs with
kinetic S-wave contribution  $\propto p$.
The P-wave nature antisymmetric under exchange of two neutrino coordinates
is dictated by that the spin part of the pair forms necessarily the symmetric triplet.
The dominance of Dirac pair emission over Majorana pair emission near thresholds
is clearly illustrated in Fig(\ref{md-asym sm2+ 2}) for the Sm$^{2+}$ target.

RANP rate is given by
\begin{eqnarray}
&&
d\Gamma_{R 2\nu} =   4 \frac{d\Gamma_R }{ \sqrt{\gamma_e^2 + \gamma_q^2} } \Gamma_{2\nu}  ( \theta) 
\,, \hspace{0.5cm}
\Gamma_{2\nu} ( \theta) = \frac{G_F^2 n}{2}  \sum_{ij}  F_{ij}  ( \theta) 
\Theta \left( {\cal M}^2 ( \theta) - (m_i +m_j )^2\right) 
\,.
\label {ranp rate}
\end{eqnarray}
In the massless neutrino limit of three flavors this becomes
\begin{eqnarray}
&&
\Gamma_{2\nu}  (\theta)
= \frac{3 G_F^2}{ 128 \pi} n  {\cal M}^2  ( \theta) 
\sim 1.1 \times 10^{-33} {\rm sec}^{-1}
 \frac{ n}{10^{15} {\rm cm}^{-3}} \, \frac{{\cal M}^2 ( \theta) }{( 0.3\, {\rm eV})^2}
\,.
\label {2nu with momentum conservation}
\end{eqnarray}
We assumed that the relevant spacing is of order 0.3\,eV.
This value is later compared with QED background rates.
One may numerically estimate  the total MC RANP rate of massless neutrino pair emission,
\begin{eqnarray}
&&
\frac{d\Gamma_{RANP} }{d\Omega } = 4
\frac{d\Gamma_{R } }{d\Omega }\frac{ \Gamma_{2 \nu} ( \theta) }{ \sqrt{\gamma_e^2 + \gamma_q^2}}
= 
\frac{27 \pi}{8} 
\frac{\gamma_{pe}  \gamma_{pq} G_F^2}{\epsilon_{pe}^2\epsilon_{pq}^2 } 
\frac{n^4 V }{\Delta \omega_0 (\gamma_e^2 + \gamma_q^2 )^{3/2} }  
\eta | \tilde{\rho}_{qg}|^2 {\cal M}^2(\theta)
\\ &&
\sim \frac{ 1.6 \times 10^{5}} {(0.1 \sim 0.5)^4 }\, {\rm sec}^{-1} \eta \frac{ 1 \,{\rm GHz} }{\Delta \nu_0} 
\frac{  {\cal M}^2  ( \theta)}{  {\rm eV}^2 }
(\frac{n }{  10^{15} {\rm cm}^{-3}})^4 \frac{V}{ 10^{-2} {\rm cm}^3} 
\,,
\end{eqnarray}
depending on a range of level spacing of $0.1 \sim 0.5$ eV
and assuming that all width factors  are  1 kHz.

The neutrino properties such as the value of smallest neutrino mass, Majorana/Dirac distinction,
CP violating phases are encoded in the angular distribution, and can be  experimentally extracted by measuring $F_{ij}(\theta)$ as
a function of Raman scattering angle $\theta$.
Angular resolution of detecting system is the most important experimentally, and the absolute rate value,
although important to confirm experimental feasibility, is not critical in deducing neutrino properties.
The resulting angular spectrum is shown in Section 4 after we discuss the background problem.

The Pauli blocking effect caused by relic neutrino of 1.9 K 
\cite{relic pauli blocking} may be included, but
we found that the spectrum distortion due to this effect is small, since
our excitation scheme does not fully use merits of an initial spatial phase imprint. 
To a good approximation for $\epsilon_{qg} \gg T_{\nu},$ this deviation is given by
$2 T_{\nu}/\epsilon_{qg}  \sim 0.87 \times 10^{-3} (100\, {\rm meV}/\epsilon_{qg}) $ 
for each species of neutrino mass eigenstate.
We shall ignore the Pauli blocking effect of relic neutrinos hereafter.

We call hereafter Raman stimulated neutrino pair emission as RANP
to distinguish it from the original scheme of RENP \cite{renp overview}.
A simplified layout of RANP experiment is illustrated in Fig(\ref{exp layout}).
To achieve a large RANP rate, it is desirable to use targets in solid environment or  ions
doped in crystals, which we shall turn to in the next section.

\vspace{0.5cm}
\section
{\bf Amplified QED backgrounds against Sm$^{2+}$ RANP in crystals}

4f valence electrons in lanthanoid ions are well protected from host crystal environment by
occupied electrons  in outer   5s and 5p shells, 
hence absorption and emission spectra of lanthanoid ions in crystals 
often exhibit sharp infrared to visible transition lines.
They seem to maintain features of  isolated ions in the free space.
Since a solid environment is required  to provide large target numbers for neutrino mass spectroscopy,
this is an ideal form of targets.
We shall consider 4f$^n$ electron system of ions in crystals
as RANP targets.

In this section we  focus on the background problem in 4f$^n$ ions in crystals.
Serious QED  backgrounds may arise when QED processes are also
macro-coherently (MC) amplified, to give rates $\propto n^2 V$ with $n$ the number
density of excited target ions, 
the case of $N$ photon emission (and absorption) being called McQN \cite{mcqn}
(MaCro-coherent Qed process of N-th order).
Rates of Raman stimulated QED backgrounds are calculable 
using the same formalism as given in the preceding section,
hence we focus on the last step ($|q \rangle \rightarrow | g \rangle $) QED process
and compare corresponding RANP rate, eq.(\ref{2nu with momentum conservation}).
From the coherence point of view it is more convenient to consider de-excitation process
$| p\rangle \rightarrow | g\rangle + \gamma + \nu_i\bar{\nu_j} $.
In this view MC amplified McQN process is
$| p\rangle \rightarrow | g\rangle + \gamma + \gamma_1 + \cdots \gamma_{N-2} $.
McQ2 process is simply the Raman scattering.
We shall show that unless McQ3 is suppressed by a device, 
the major QED background is much stronger than RANP, hence
shall discuss  how to get rid of this background.
It turns out that McQ3 background of $\gamma_0 + | e\rangle \rightarrow | g \rangle + \gamma + \gamma' $ 
should be, and can be, completely rejected by a choice of kinetics and
the next dangerous McQ4 background should be forbidden by a symmetry principle of host crystals.
Still higher QED backgrounds have negligible rates.

Note first that excitation by  two laser irradiation creates a spatial phase factor
 $e^{i\vec{p}_{eg}\cdot\vec{x}}  $
imprinted to ion targets, which have a momentum $ \vec{p}_{eg}$ 
as if  given to the initial state $|e\rangle$.
Its magnitude in two-photon excitation scheme depends on how two lasers are irradiated:
when they are irradiated from the same direction, $|\vec{p}_{eg}| = \epsilon_{eg}$ irrespective of
whether it is of ladder-type or  of Raman-type excitation.
When they are irradiated from the counter-propagating directions,
$|\vec{p}_{eg}| = r \epsilon_{eg}$ with $r \leq 1$ for ladder-type excitation and  $ r \geq 1$ for Raman-type excitation.
We consider excitation by two lasers from the same direction.

A criterion of how QED backgrounds appear in the angular distribution
 is given by using the squared mass  function 
$ {\cal M}^2(\theta)$ defined in eq.(\ref{squared mass f}).
When the most dangerous McQ3 background exists,
 this quantity coincides with the zero squared mass of photon at some angle,  
thus $ {\cal M}^2(\theta) $ vanishes
at an angle $\theta$ in  $-\pi < \theta \leq \pi $.
Hence the condition of McQ3 rejection is that this quantity is positive definite for any $\theta$.
In RANP this quantity is equal to the neutrino pair mass 
$(p_1 + p_2)^2 = (E_1+E_2)^2 - (\vec{p}_1 + \vec{p}_2 )^2$,
which is larger than $4 m_1^2$ with $m_1$ the smallest neutrino mass of three neutrinos.
The  squared mass function for Sm$^{2+}$ case is illustrated
 against the Raman scattering angle $\theta$ for a few choices of the trigger
angle $\theta_0$ in Fig(\ref{mcq3 sm2+}).
It is found in this case that McQ3 rejection is possible  for $\theta_0$ chosen to be close to $\pi$.

Having rejected the McQ3 background, the next problem is the McQ4 background.
Hence we consider $| q\rangle \rightarrow | g\rangle + \gamma_1 + \gamma_2 $ where
$\gamma_i\,,  i = 1,2$ are additional emitted  photons besides the Raman scattered photon.
In this case the squared mass function ${\cal M}^2$ is equal to 
Lorentz invariant $(k_1 + k_2)^2 = 4\omega_1 \omega_2 \sin^2(\theta_{12}/2)$ of two-photon pair,
and it is difficult to kinetically reject this background: 
although the photon is massless, this function $ (k_1 + k_2)^2$ can take any non-negative value including zero.

How large is the expected McQ4 background ?
We shall estimate this rate by calculating the two-photon emission rate at $ | q\rangle \rightarrow | g\rangle $
and comparing to the corresponding the neutrino pair emission rate.
The expected major background is of M1 $\times $ E1 type two-photon emission (see below on more of this).
MC two-photon total emission rate of  M1 $\times $ E1 at $| q\rangle \rightarrow | g\rangle $ is given by
\begin{eqnarray}
&&
\Gamma_{2\gamma}(\theta) = \frac{9\pi}{8} \frac{n}{\sqrt{\epsilon_{qg}^2 - {\cal M}^2(\theta)}}
\sum_a \,\frac{ \gamma_{aq}  \gamma_{ag} }{ \epsilon_{aq}^2 \epsilon_{ag}^2} 
\int_{\omega_-}^{\omega_+} d\omega \, \omega (\epsilon_{qg} - \omega )
\left( \frac{1}{\epsilon_{aq} + \omega} + \frac{1}{\epsilon_{ag} - \omega}
\right)^2
\,, 
\\ &&
\omega_{\pm} = \frac{1}{2} (\epsilon_{qg} \pm  \sqrt{\epsilon_{qg}^2 - {\cal M}^2(\theta)})
\,.
\end{eqnarray}
This rate is to be compared with the corresponding MC amplified neutrino pair emission rate $\Gamma_{2\nu}(\theta)$
of eq.(\ref{2nu with momentum conservation}).
By taking an intermediate state $| a\rangle $ far above $| q\rangle $,
$ \epsilon_{aq} \gg \epsilon_{qg}$, one can  derive a lower limit of the ratio, to give
\begin{eqnarray}
&&
\frac{\Gamma_{2\gamma}(\theta)  }{ \Gamma_{2\nu}(\theta)} > 
 \frac{4 \pi^2 \gamma_{aq}  \gamma_{ag} }{G_F^2 \epsilon_{aq}^2 \epsilon_{ag}^2 }
\left( \frac{1}{\epsilon_{aq} } + \frac{1}{\epsilon_{ag}} \right)^2
\frac{{\cal M}^2(\theta) + 2\epsilon_{qg}^2 }{ {\cal M}^2(\theta)} >
\frac{4 \pi^2 \gamma_{aq}  \gamma_{ag} }{G_F^2 \epsilon_{aq}^2 \epsilon_{ag}^2 }
\left( \frac{1}{\epsilon_{aq} } + \frac{1}{\epsilon_{ag}} \right)^2
\,.
\end{eqnarray}
4f lanthanoid system suggests that decay rates are of order 1/msec,
and energy spacings are of order $0.5 \sim 2 $eV.
These typical values give this ratio much larger than $10^{21}$.

We shall analyze the McQ4 background problem in more detail by taking the concrete case of lanthanoid ions.
Since 4f electrons are insensitive to host crystal environment,
we may approximate ion wave functions based on the standard Russel-Sanders scheme using the
$^{2S+1}L_J $ notation in the free space \cite{atomic physics text}, 
and introducing their small mixture at the next stage of approximation.
This allows one to use the concept of angular momentum valid in the free space, 
which is  modified by  crystal field effects, as discussed below.
An important constraint on optical transitions among 4f$^n$ manifolds arises from time reversal  (T-reversal) symmetry
which holds strictly in crystals when no external magnetic field is applied, which we assume
hereafter.
We amplify rates by generating coherence between
two 4f$^n$ states of $|p \rangle $ and $|g \rangle $ which, we assume, are electromagnetically connected
by T-reversal even  operators (there may be another choice, but for definiteness we consider this case).
We further assume that the relative quantum number of two states
 at $|p \rangle \rightarrow |q \rangle $ is T-reversal odd.
The constraint from time reversal symmetry then restricts the major
 QED two-photon background at $|q \rangle \rightarrow |g \rangle $ to be of  type  M1 (magnetic dipole)
$\times $ E1 (electric dipole) type transition, and the next major to
M1 $\times $ E2 (electric quadrupole) type transition bypassing a larger M1 $\times$ M1.

One needs symmetry to forbid this large background.
Before we discuss its possibility, we mention that there are two aspects of
McQ4 background rejection.
Most of McQ4 backgrounds emit detectable extra photons in addition to one Raman scattering photon
in RANP case.
Thus, one can directly observe McQ4 events and subtract this contribution from data.
Dangerous events are those of McQ4 events in which both of two extra photons escape detection.
The number of events of this class may be small, but one needs dedicated simulation
of these missing events.
Another concern is that the occupied number of ions in $|q \rangle$
might be depleted almost completely.
In particle physics terminology this is the problem of the small branching ratio.
For instance, a few GHz  estimated for absolute  RANP event rates might be
actually a few Hz  if the branching rate  $10^{-9}$ is taken into account.

We now discuss the possibility of using crystal symmetry to forbid M1 $\times $ E1 two-photon emission
at  $|q \rangle \rightarrow | g \rangle$.
Even between two 4f$^n$ levels, E1 transitions may occur roughly with comparable rates to
M1 in host crystals of lower symmetry without inversion center, 
as pointed out by \cite{van vleck}.
There are however cases in which lanthanoid ions are located at inversion center of highly symmetric
host crystal.
A number of crystals having inversion center are limited, 10 out of 32 crystal point groups 
\cite{inui et al textbook}.
Furthermore, dopants may not be at the inversion center even if they are doped in crystals of  10 groups.
Fortunately, we found by looking at crystal structures of possible host crystals that
alkali-earth halide crystals such as CaF$_2$ and SrF$_2$ are promising hosts, having the point group symmetry O$_h$ \cite{inui et al textbook}, which is known to preserve parity at alkali-earth ion sites.
Matched lanthanoid dopant ion is divalent instead of more popular trivalent ions when ions substitute
Ca$^{2+}$ ion in alkali-earth halides \cite{rubio}.
These crystals have been studied since the early day's of the search history of lasing solids.
Another idea is to use O$_h$ symmetric crystals such as SmF$_2$, SmH$_2$ directly  for the divalent ion.
It is necessary in this case to verify that relaxation processes associated with phonons
are well suppressed.

The next major MC amplified QED process, M1 $\times $ E2, is neither tolerable,
because its rate is only of order $\alpha^2$ smaller.
Our proposal to further forbid McQ4 M1 $\times $ E2 emission is to choose
the angular momentum selection rule such that this process is forbidden by
$|\Delta J| \geq 4$ between $| q\rangle $ and $| g\rangle $, since  M1 $\times $ E2
two-photon transition changes the angular momentum by $|\Delta J| \leq 3$.
The selection rule based on the angular momentum conservation is strictly valid only
in the free space, and we now have to consider effects of crystal field.
According to \cite{van vleck},
there exists dynamical effect of enhanced electric multi-pole transitions including E1 that may occur
even if the system allows a point group symmetry having inversion center.
This arises from a coupled hamiltonian term of dipole and lattice vibration
derived from crystal field potential.
The mechanism was formulated in the case of lower symmetry by \cite{judd-ofelt},
and in a high symmetry case by \cite{stark split oh}, \cite{ho2+ forced e1}.
We shall explain the mechanism, taking our example.

\begin{figure*}[htbp]
 \begin{center}
 \epsfxsize=0.5\textwidth
 \centerline{\includegraphics[width=10cm,keepaspectratio]{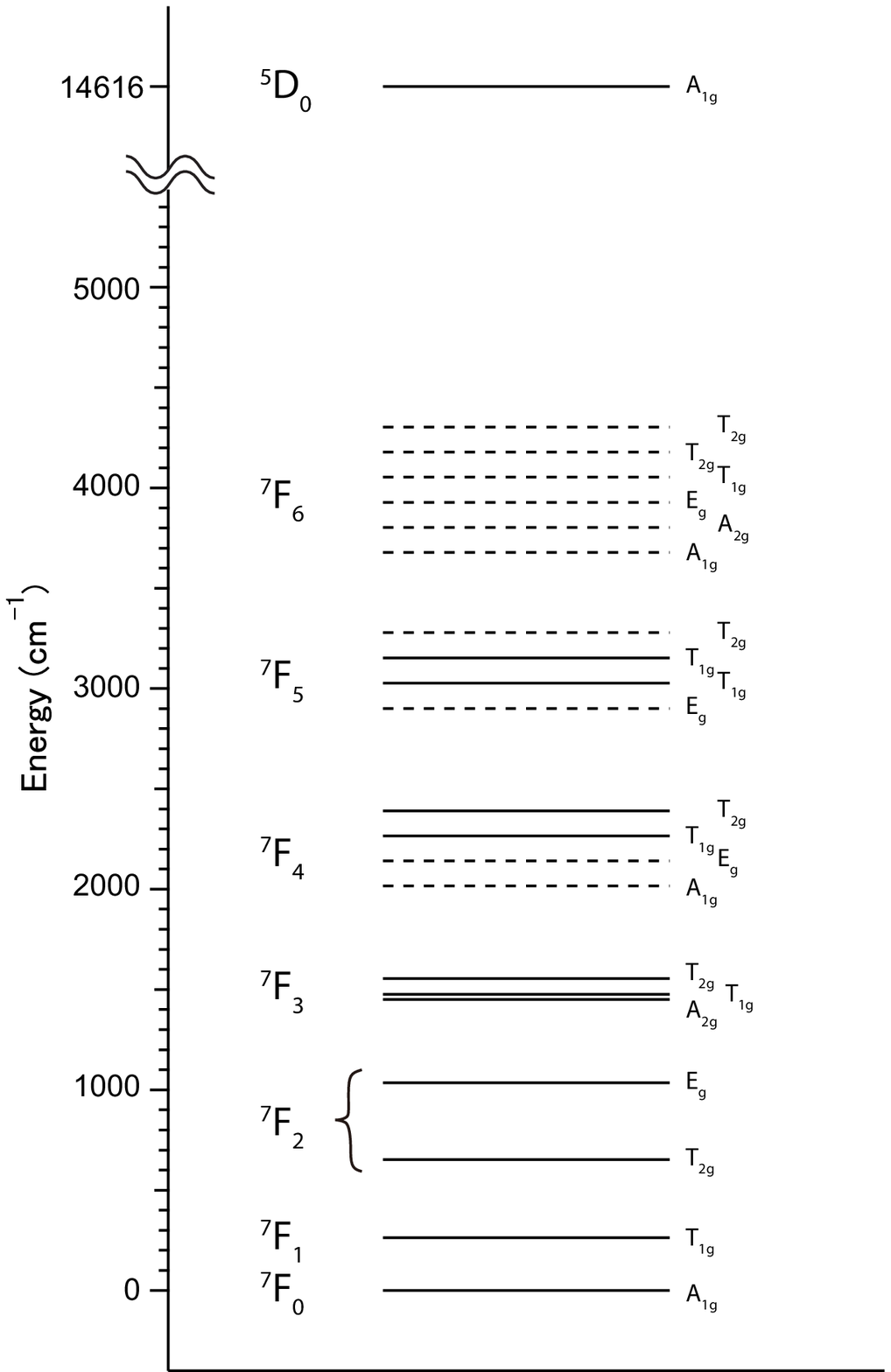}} \hspace*{\fill}
   \caption{
Sm$^{2+}$: SrF$_2$ level structure.
Solid lines are for optically identified levels, and dashed lines for theoretically calculated ones 
by \cite{sm2+  data}.
}
   \label {sm2+ levels}
 \end{center} 
\end{figure*}

Consider 4f electron of divalent lanthanoid ion Sm$^{2+}$ at an inversion center of alkali-earth halides
with O$_h$ symmetry.
It is useful to keep in mind the level structure of this ion shown in Fig(\ref{sm2+ levels})
for the following discussion.
Adding the Coulomb field  of  nearest eight fluoride ions,
the crystal field $V_C$ at the inversion center is given by
\begin{eqnarray}
&&
V_C =  - c \frac{Z \alpha }{a^5}
\left(x^4 + y^4 + z^4 - \frac{3}{5} r^4 +  O (x_i^6)
\right)
\,, \hspace{0.5cm}
c Z = \frac{2^7 \cdot 5 \cdot 7}{3^5}\frac{Z}{2} \simeq 18.436 \frac{Z}{2}
\,,
\label {crystal field expansion}
\end{eqnarray}
with $\alpha \simeq 1/137$ the fine structure constant.
$Z=2$ for Sm$^{2+}$:SrF$_2$.
The crystal field thus given, or its extension including more surrounding ions and
departure of the point source model, is totally responsible for Stark splitting among components of J-manifolds.
We assume that this is the only source of dopant energy shift and other dynamical effects
in crystals.
The Coulomb energy at the lattice constant gives a scale of this hamiltonian,
$\alpha/a = 2.5\, {\rm eV} (5.8 \times 10^{-8} {\rm cm}/a) $.
The lattice constant of SrF$_2$ is $\sim 5.8 \times 10^{-8} {\rm cm}$.
The important constraint from crystal symmetry is that the crystal field $V_C$ of eq.(\ref {crystal field expansion})
belongs to a singlet irreducible representation A$_{1g}$ under O$_h$  \cite{inui et al textbook}.
The electron's coordinate depends on a lattice point $\vec{l}$,
its deviation from equilibrium point by lattice vibrating coordinate $\vec{Q}$ like
\begin{eqnarray}
&&
\vec{r} = \vec{l} + \vec{Q} + \vec{r}\,' 
\,.
\end{eqnarray}
For simplicity we use a shorthand notation $\vec{r} $ for $\vec{r}\,' $ below.
When this is inserted into expansion of crystal field $V_C$, it contains vibration-electronic interaction
via operators, $rrr Q \,,  r r QQ\,, r Q QQ$ in the order of strength.
The lattice vibration $Q$  may be expanded in terms of phonon annihilation $a_s $ and 
creation $a_s^{\dagger} $ operators of normal modes:
\begin{eqnarray}
&&
\vec{Q} = \frac{1}{\sqrt{ 2M N}} \sum_s  \frac{\vec{e}_s}{\sqrt{\omega_s }} \left(a_s e^{i \vec{q}\cdot \vec{l} } + 
a_s^{\dagger} e^{ - i \vec{q}\cdot \vec{l} } 
\right)
\,,
\label {lattice vibration op}
\end{eqnarray}
with $s = (\vec{q}\,, \vec{e}_s)$ and $M$ the ion mass of $\sim 1.5 \times 10^{11}$eV for the ion of mass
number 150 (an isotope of stable Sm).
$N$ is the total number of ions in a crystal.
The phonon polarization vector is normalized as
$\vec{e}_s\,^{*} \vec{e}_{s'}  = \delta_{s s'}$.
The presence of a large nuclear mass $M$ makes expansion in terms of emitted phonon numbers
a useful concept.

\begin{figure*}[htbp]
 \begin{center}
 \epsfxsize=0.6\textwidth
 \centerline{\includegraphics[width=17cm,keepaspectratio]{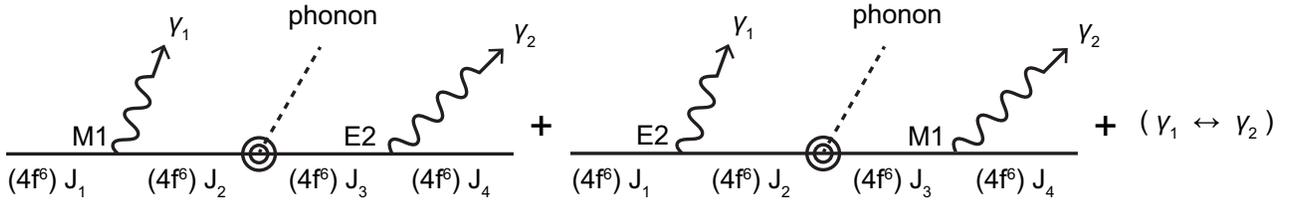}} \hspace*{\fill}
   \caption{
Perturbative Feynman diagrams for phonon-assisted, two-photon emission in lanthanoid ion transition
of manifolds,  $J_1 \rightarrow J_4$.
Dashed lines are for phonon and wavy lines for photons,
double circles indicating three electron position $r_i r_j r_k$ operators inserted at a phonon emission.
}
   \label {background diagram}
 \end{center} 
\end{figure*}

We first discuss the most important background arising from $rrr Q$ term, and
may call this process a single-phonon assisted M1 $\times$ E2 two-photon emission in crystals.
One can follow angular momentum changes at three vertexes in the diagram.
In lanthanoid ion system we shall discuss in the next section, the last path 
$|q \rangle \rightarrow | g\rangle $
involves a large angular momentum, either $-5 $ or $-4 $. 
The maximum change of angular momentum in $rrr Q$ 
is associated with the following expanded term,
\begin{eqnarray}
&&
\Delta V_C^{(Q)} = - 2c \frac{ \alpha }{a^5}
\left( r_-^3 (Q_x + i Q_y ) + ({\rm h.c.} ) \cdots
\right)
\,.
\end{eqnarray}
The electron operator $r_-^3 $ lowers the angular momentum by three units
When this operator is inserted in the middle of M1 and E2 QED vertexes as in Fig(\ref{background diagram}),
the combined angular momentum change can be matched without a conflict.

The effect of crystal field insertion may  be discussed separately from QED  M1 $\times$ E2 two-photon emission
part: indeed the QED part coincide with that of the free space expression.
The probability amplitude of  M1 $\times $ E2 two-photon de-excitation in the free space is worked out
in a standard way using the second order of perturbation theory, to give
\begin{eqnarray}
&&
\hspace*{-0.5cm}
{\cal A}_{2\gamma}( \omega_1, \omega_2) =
\frac{ e^2}{ 4m_e^2 } \sum_n \left( \frac{\omega_2^2\, \vec{r}_{ng} \cdot\vec{A}^{(2)} 
\langle J_y + S_y \rangle_{ng}B_y^{(1)} }
{ \epsilon_{nq} + \omega_1 } + 
\frac{ B_y^{(2)} \langle J_y + S_y\rangle_{ ng}
   \omega_1^2\, \vec{r}_{nq} \cdot\vec{A}^{(1)} } 
{ \epsilon_{ng} + \omega_2 } + ( 1 \leftrightarrow 2 )
\right) 
\,, 
\label {m1 x e2 amplitude}
\end{eqnarray}
where we took E2 emitted photon along z-direction with a linear polarization along x-axis.
$B_y^{(i)} ( \vec{A}^{(i )}) $ is a magnetic field (vector potential) of emitted photon.
With $\vec{A} = e^{i \vec{k}\cdot \vec{x }} \vec{\epsilon} /\sqrt{ 2 \omega V_q}\,, V_q=$ the quantization volume,
${\cal A}_{2\gamma}( \omega_1, \omega_2) $ has a correct mass dimension of energy, +1.

Most important intermediate states $|n \rangle $ that  dominantly 
contribute to M1 $\times$ E2 two-photon emission
come from levels of lower energies than $|q \rangle $ if there are any,
possibly via resonance.
We shall discuss the resonance problem later, and here give a simple order of magnitude estimate.
We approximate the energy denominator in eq.(\ref{m1 x e2 amplitude}) by a typical constant value
 $ \bar{\epsilon} $:
\begin{eqnarray}
&&
{\cal A}_{2\gamma}( \omega_1, \omega_2) \sim 
\frac{ e^2  \omega_1^2 }{ 4m_e^2 \bar{\epsilon} } 
\left(  B_y^{(2)} \langle J_y + S_y\rangle_{ ng}
  \, \vec{r}_{nq} \cdot\vec{A}^{(1)} \right) + ( 1 \leftrightarrow 2 )
\,.
\end{eqnarray}
The total probability amplitude of a phonon-assisted, two-photon emission is then
\begin{eqnarray}
&&
{\cal M}_{2\gamma}^{(Q)} =  {\cal A}_{2\gamma}( \omega_1, \omega_2)  \frac{\Delta V_C^{(Q)} }{ \bar{\epsilon} }
\,.
\end{eqnarray}
The rest of rate calculation is standard in particle physics, using
\begin{eqnarray}
&&
\Gamma_{2\gamma}^{(Q)} =
\frac{n}{ (2\pi)^5} V_q^3 \,\int  d^3 k_1 d^3 k_2 \int   d^3 q\,
\delta (\omega_1 + \omega_2 + \omega_s(q) - \epsilon_{qg} ) \delta (\vec{k}_1 + \vec{k}_2 + \vec{q} - \vec{p}_{eg} - \vec{k}_0 ) 
\, |  {\cal M}_{2\gamma}^{(Q)}|^2
\\ &&
\hspace*{1cm}
\equiv \int d{\cal P} V_q^3  |  {\cal M}_{2\gamma}^{(Q)}|^2
\,, \hspace{0.5cm}
V_q^3  |  {\cal M}_{2\gamma}^{(Q)}|^2 = 0.036 \alpha^2 
\frac{\omega_1 \omega_2 ( \omega_1 + \omega_2)^2\, r_{4f}^8} { m_e^4 \bar{\epsilon}^4 a^{10} M n
\omega_s(q)}
\,.
\end{eqnarray}
As expected, the quantization volume $V_q $ dependence disappears in the final result, since three emitted particles, two photons
and a phonon squared wave functions have $1/V_q^3$ with $N = n V_q$ in the phonon expansion
is considered.
We shall take the case of acoustic phonon for which $\omega_s(q) = c_1 q\,, c_1 = O(10^{-5})$.
The phase space integral in this approximation  gives a simple analytic result,
\begin{eqnarray}
&&
\int d^3 k_1 \int d^3 k_2 
\frac{ \omega_1 \omega_2 ( \omega_1 + \omega_2)^2 }{ \epsilon_{qg} - \omega_1- \omega_2 }
\delta (\omega_1 + \omega_2 + \omega_s(q) - \epsilon_{qg} ) 
= \frac{ (2\pi)^2 \epsilon_{qg}^9 }{ 270 c_1^2  |\vec{p}_{eg} + \vec{k}_0|}
\,.
\end{eqnarray}
The final result of total phonon-assisted, two-photon emission rate is given by
\begin{eqnarray}
&& 
\Gamma_{2\gamma}^{(Q)} =  (\frac{ 2c \pi}{ \sqrt{2} \,6^3} )^2 
\frac{\alpha^4 r_{4f}^8 \langle J+S \rangle^2} {m_e^4 \bar{\epsilon}^4 a^{10} M c_1^2 }  
\frac{ \epsilon_{qg}^9 }{ 270  |\vec{p}_{eg} + \vec{k}_0|}
\,.
\end{eqnarray}
Electron position matrix elements, $r_{ni}$, that appear in the original amplitude (\ref{m1 x e2 amplitude}),
were replaced by typical size of 4f electron for a simple estimate.
Numerically, the total rate is of order,
\begin{eqnarray}
&&
\Gamma_{2\gamma}^{(Q)} \sim 2 \times 10^{- 36} \, {\rm sec}^{-1}\,
 \frac{  \epsilon_{qg}^9 }{  |\vec{p}_{eg} + \vec{k}_0|}\,({\rm0.3  eV})^{-8}
\,.
\end{eqnarray}
We took $r_{4f} = 0.25 \times 10^{-8}{\rm cm}\,, a= 5.8  \times 10^{-8}{\rm cm}\,, \bar{\epsilon}= 1{\rm eV} \,,
c_1 = 10^{-5}$.
This rate for McQ4 background in crystals should be compared to the corresponding rate
$1 \times 10^{-33} {\rm sec}^{-1}  $ of eq.(\ref{2nu with momentum conservation}) for RANP.

The optical phonon emission rate instead of acoustic phonon 
can be estimated too, to give a phase space integral.
$(\epsilon_{qg} - \omega_{op})^9/35\,  \omega_{op}  $,
replacing $\epsilon_{qg}^9/( 270  c_1^2 |\vec{p}_{eg} + \vec{k}_0|) $ in the acoustic phonon case.
The optical phonon frequency at zero momentum $ \omega_{op} $ is of order 10 meV,
and it can be shown that the optical phonon emission rate is far below the acoustic phonon emission.

Even if this small background rate is an overestimate,
 McQ4 backgrounds produce
two extra photons besides a single Raman scattered photon in RANP case and 
its experimental identification and isolation should not be a problem.

Let us estimate the next leading contribution, arising from crystal field expansion,
eq.(\ref{crystal field expansion}), of the form,
$r_i r_j Q_k Q_l$.
The ratio of  rate arising from this term to the single phonon emission rate already discussed is of order,
$( Q/r)^2$, which is estimated as
\begin{eqnarray}
&&
(\frac{Q}{r})^2 = \frac{ 1}{2 M n V\, \omega_s(q) \, r^2 } \sim 2 \times 10^{-17 }
\,,
\end{eqnarray}
taking typical parameters.
Hence, contributions from this and higher order expansion terms  are negligible as  backgrounds
against RANP.

One may well wonder what happened to the angular momentum change in
the case of neutrino-pair emission.
The answer is that
Raman stimulation imparts to ions not only the linear momentum but also the
angular momentum. The angular momentum change can be understood by 
inspecting the spatial phase factor,
$e^{i (\vec{P} - \vec{p}_1 - \vec{p}_2)\cdot \vec{r}}\,,  \vec{P} = \vec{p}_{eg} + \vec{k}_0 - \vec{k}$,
which led to the macro-coherence condition,
namely, the  momentum conservation $\vec{P} - \vec{p}_1 - \vec{p}_2 = 0 $.
Thus, the neutrino-pair carries away the momentum of finite amount $\vec{P} \neq 0$
along with the angular momentum larger than 4,5 when Raman scattering  occurs at angles
away from pair thresholds.

\vspace{0.5cm}
\section
 {\bf Photon angular distribution in Sm$^{2+}$ doped crystal RANP}

The most popular lanthanoid ions doped in dielectric crystals are trivalent ions.
In popular host crystals such as YLF and YAG these trivalent ions are
at sites without inversion symmetry.
For instance, in the YLF case the symmetry at the site of trivalent ions
is S$_4$ which does not have inversion center \cite{trivalent ion location}, 
\cite{trivalent ion doped crystals}.
We find no good trivalent ion candidate.

Our idea on  host crystals is to use  O$_h$ symmetric alkaline-earth halides such
as SrF$_2$ and CaF$_2$, and to  substitute divalent alkaline-earth ions, Sr$^{2+}$
and Ca$^{2+}$,  by  divalent lanthanoid ions at the inversion center of O$_h$ symmetry.
Candidate divalent lanthanoid ions have to be carefully selected 
to eliminate a remaining background possibility, M1 $\times $ E2 two-photon decay
at $|q \rangle \rightarrow | g\rangle $.
The simplest way to minimally reduce this  is to use the angular momentum selection rule
(although approximate) of
$| \Delta J| \geq 4$ since  M1 $\times $ E2 requires $| \Delta J| \leq 3$ \cite{violated angular momentum}.
Existence of many J-manifolds is essential to realize this idea, and we are led to
Sm$^{2+}$ 4f$^6$ system as a good candidate ion.
Sm$^{2+}$ is incidentally the divalent lanthanoid ion most extensively studied.

Sm$^{2+}$ 4f$^6$ system has eight J-manifolds, 
$^7$F$_0$,  $^7$F$_1$,  $^7$F$_2$,  $^7$F$_3$,  $^7$F$_4$, 
 $^7$F$_5$,  $^7$F$_6$, $^5$D$_0$
in the notation  $^{2S+1}L_{J}$.
The quantum number assignment based on O$_h$ symmetry
 is shown along with level spacings in Fig(\ref{sm2+ levels}).
Some levels are optically identified, and others are not.
These J-manifolds are split by crystal field  into irreducible representations of crystal symmetry: 
triplets T$_{1g}$, T$_{2g}$, doublet E$_g$, and singlet A$_{1g}$.
 Around 0.5 eV above the  $^7$F$_6$ manifold,  $^5$D$_0$ manifold is at $\sim$ 1.8 eV from the ground state.
Data of energy levels in alkali-earth hallides are given in \cite{sm2+  data} along with some optical information.
We can think of several RANP paths to select initial and intermediate states:
$|e \rangle $, $|p \rangle $, $|q \rangle $.
Unfortunately, T-reversal quantum numbers of these states are not known at present.

We have examined two promising path schemes, but from the point of McQ3 background rejection
it turned out that only one of them is acceptable.
Optimized Sm$^{2+}$: SrF$_2$ RANP scheme 
are then as follows.
We adopt inelastic Raman stimulation scheme of $\epsilon_{pe}  > \epsilon_{pq}$,

$|e \rangle $: $^7$F$_4$, T$_{1g}$ 2266 cm$^{-1}$ (280.5 meV),

$|p \rangle $: $^7$F$_6$, T$_{1g}$  4053 cm$^{-1}$ (501.7 meV),

$|q \rangle $: $^7$F$_4$, T$_{2g}$ 2391 cm$^{-1}$ (296.0 meV),

$|g \rangle $: $^7$F$_0$, A$_{1g}$ 0 cm$^{-1}$.

A part of level is not optically identified in the host of SrF$_2$,
but suggested by crystal field calculation  \cite{sm2+  data}.
The state $^7$F$_6$, T$_{1g}$ is introduced by theoretical calculation of Stark levels whose
parameters are derived from optical data related to confirmed levels.

\begin{figure*}[htbp]
 \begin{center}
 \epsfxsize=0.6\textwidth
 \centerline{\includegraphics{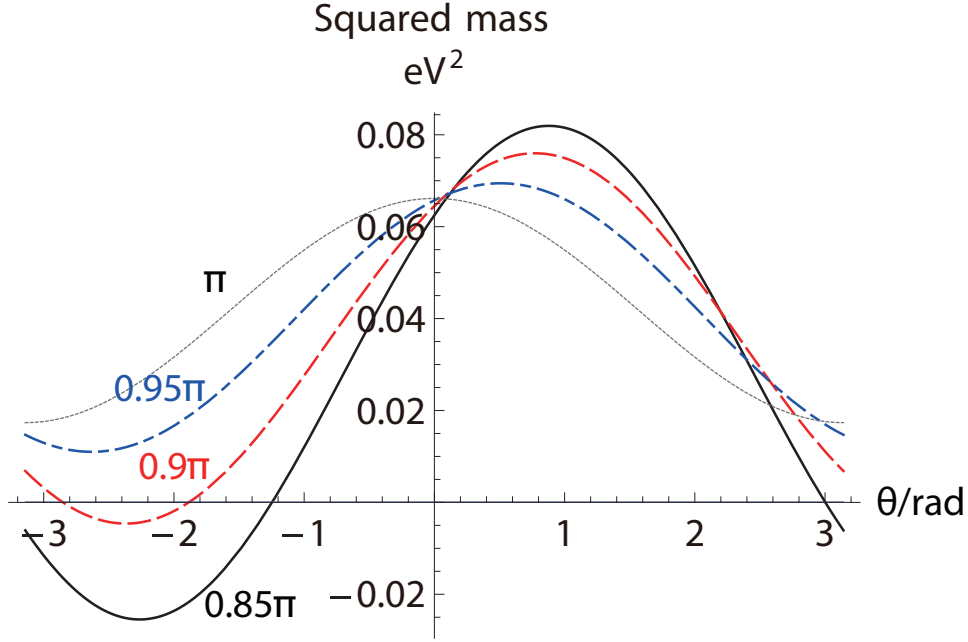}} \hspace *{\fill}
   \caption{
Inelastic Raman ${\cal M}^2 (\theta)$ distribution of Sm$^{2+}$ scheme 
against Raman scattered angle $\theta$, taking a few choices of Raman trigger direction $\theta_0$. 
$\theta_0 = 0.85 \pi$ in solid black, $0.9 \pi$ in dashed red, $0.95 \pi$ in dash-dotted blue,
 and $\pi$ in dotted black.
}
   \label {mcq3 sm2+}
 \end{center} 
\end{figure*}

The squared mass given to neutrino pairs is
\begin{eqnarray*}
&&
{\cal M}^2(\theta) 
= 2 \epsilon_{eg}\left( \epsilon_{pe} (1- \cos \theta_0) -  \epsilon_{pq} (1-\cos\theta)\right) 
- 4 \epsilon_{pe} \epsilon_{pq} \sin^2 \frac{\theta - \theta_0}{2}
\,,
\end{eqnarray*}
with $|\vec{p}_{eg} | =  \epsilon_{eg}$ and $\epsilon_{pe}  = 221\, {\rm meV}\,, \epsilon_{pq} = 205.7 \, {\rm meV}$.
The condition of McQ3 rejection for the specified path is given by ${\cal M}^2(\theta) > 0 $ for any $\theta$.
In Fig(\ref {mcq3 sm2+}) $\sim $ Fig(\ref{ang rate sm2+ 22}) we illustrate how
McQ3 rejection is made possible and an example of resulting Raman angular spectrum.

\begin{figure*}[htbp]
 \begin{center}
 \epsfxsize=0.6\textwidth
 \centerline{\includegraphics{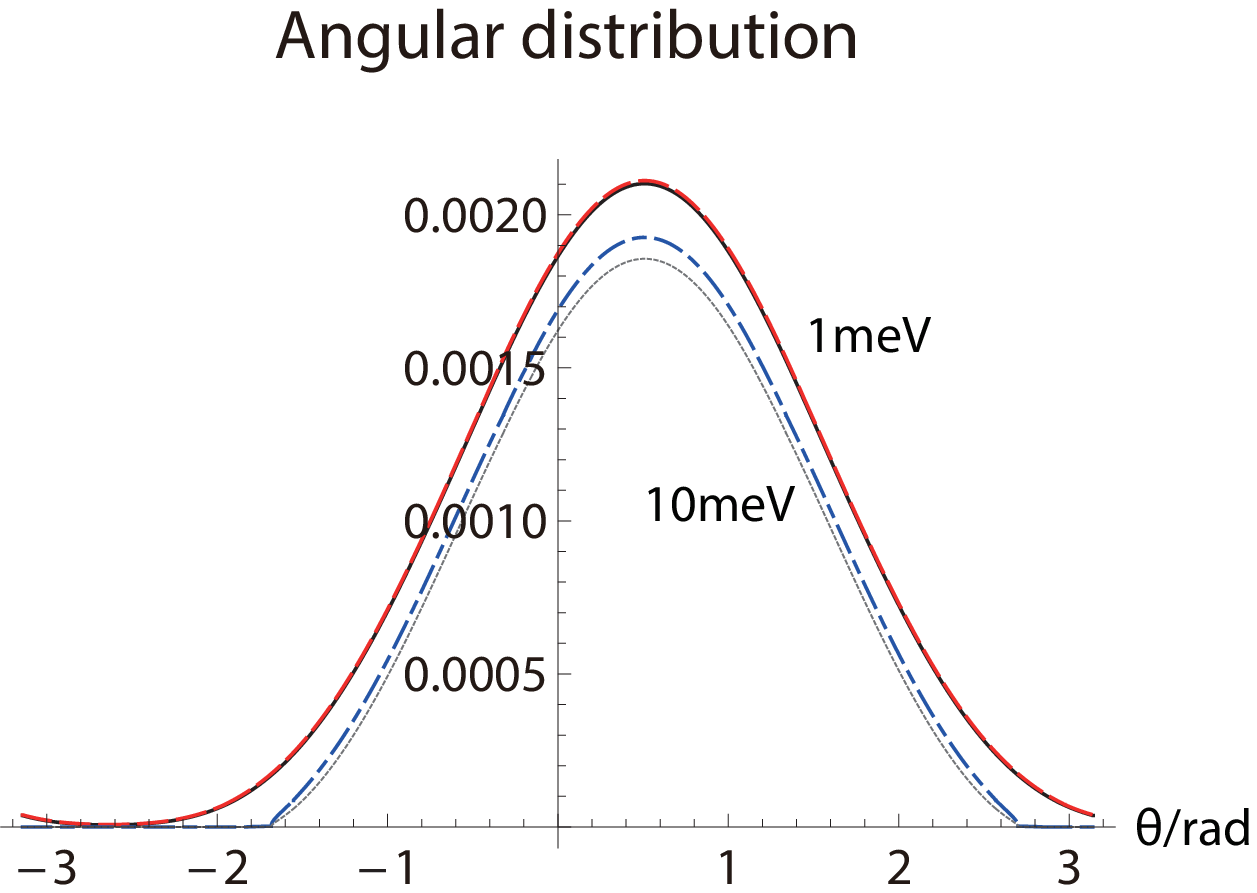}} \hspace *{\fill}
   \caption{
Photon angular distribution of Sm$^{2+}$ scheme, 
taking the trigger Raman angle $\theta_0 = 0.95 \,\pi $ (giving events without McQ3 background)
Majorana NH case of smallest neutrino mass 1 meV in solid black, 10 meV in dotted black,
Dirac NH case of 1 meV in dashed red, and 10 meV in dash-dotted blue.
The absolute rate to be multiplied to numbers here
 is  $6 \times 10^{10} $sec$^{-1} \eta  | \tilde{\rho}_{qg}|^2 $, taking $n= 10^{16}\,{\rm cm}^{-3}\,,
V = 0.01 \,{\rm cm}^{3}$.
}
   \label {ang rate sm2+ 2}
 \end{center} 
\end{figure*}

\begin{figure*}[htbp]
 \begin{center}
 \epsfxsize=0.6\textwidth
 \centerline{\includegraphics{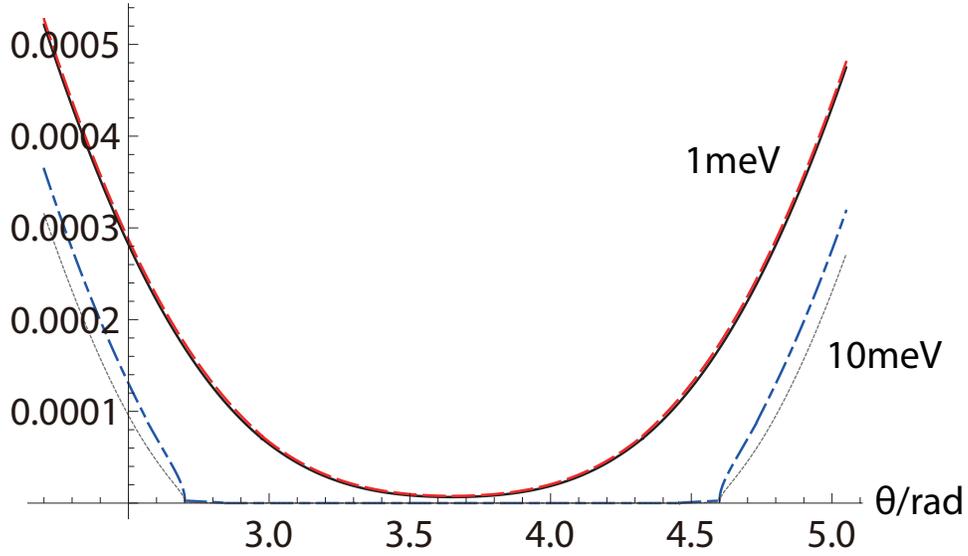}} \hspace *{\fill}
   \caption{
Enlarged threshold region  of Fig(\ref{ang rate sm2+ 2}), assuming the same parameter set
as Fig(\ref {ang rate sm2+ 2}).
}
   \label{ang rate sm2+ 22} 
 \end{center} 
\end{figure*}

\begin{figure*}[htbp]
 \begin{center}
 \epsfxsize=0.6\textwidth
 \centerline{\includegraphics[width=12cm,keepaspectratio]{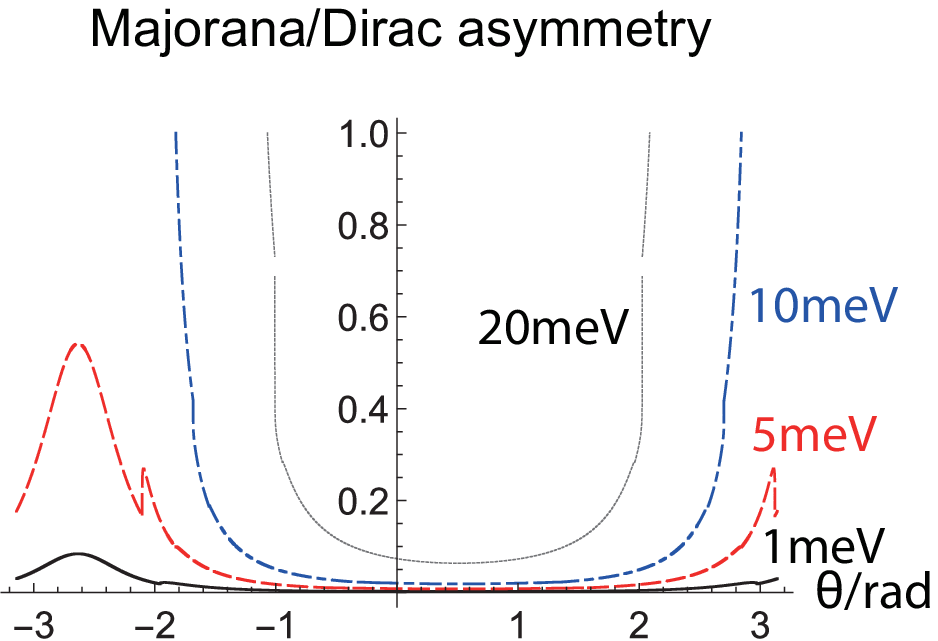}} \hspace *{\fill}
   \caption{
Majorana/Dirac asymmetry
given by (Dirac rate - Majorana rate)/(Dirac rate + Majorana rate) with $\theta_0=0.95\pi$.
NH smallest neutrino mass 1 meV in solid black, 5 meV in dashed red,  10 meV in dash-dotted
blue, and 20 meV in dotted black.
}
   \label {md-asym sm2+ 2}
 \end{center} 
\end{figure*}

From these figures we find that the mass measurement is relatively easier down to the level
of smallest neutrino mass of order 1 meV.
The Majorana/Dirac asymmetry may be defined by
(Dirac rate - Majorana rate)/( Dirac rate + Majorana rate).
The asymmetry thus defined is plotted in Fig(\ref{md-asym sm2+ 2}) for Sm$^{2+}$ scheme.
If the smallest neutrino mass is found less than a few meV,
the Majorana/Dirac distinction may require a high statistics data.

We now discuss more details of de-excitation,  excitation and coherence generation schemes 
along with Raman trigger in $|e\rangle \rightarrow |p\rangle$ based on the point group symmetry
O$_h$.
First we note that Sm$^{2+}$ de-excitation scheme uses states belonging to
irreducible representation (irrep) of O$_h$ symmetry:
\begin{eqnarray}
&&
{\rm T}_{1g}^- \rightarrow {\rm T}_{1g}^+ \rightarrow {\rm T}_{2g}^- \rightarrow {\rm A}_{1g}^+
\,.
\end{eqnarray}
We took an example of T-reversal quantum numbers denoted here by $\pm$ for definiteness.
Within Sm$^{2+}$ 4f$^6$ manifolds the dominant transition is of magnetic dipole M1$^-$ type,
which belongs to irrep $\tilde{{\rm T}}_{1g}^-$.
We distinguish irreducible representation of operator by putting tilde $\tilde{}$.
Excitation $|g\,, {\rm A}_{1g}^+ \rangle \rightarrow |e\,, {\rm T}_{1g}^- \rangle$ is possible by
irradiating two lasers  which induce M1$^- \times$ E2$^+$ 
(E2$^+$ belonging to $\tilde{{\rm T}}_{2g}^+ +  \tilde{ {\rm E}}_g^+$, triplet + doublet) 
transition due to product decomposition
of two irreps:
\begin{eqnarray}
&&
{\rm A}_{1g}^+ \times \tilde{{\rm T}}_{1g}^- = {\rm T}_{1g}^-
\,, \hspace{0.5cm}
{\rm T}_{1g}^- \times \tilde{{\rm T}}_{2g}^+ =  {\rm T}_{1g}^- + {\rm T}_{2g}^- + {\rm E}_g^- + {\rm A}_{2g}^-
\,, \hspace{0.5cm}
{\rm T}_{1g}^- \times \tilde{ {\rm E}}_g^+ =  {\rm T}_{1g}^- + {\rm T}_{2g}^-
\,.
\end{eqnarray}

One can think of Raman-type of excitation  from the same irradiation direction
for Sm$^{2+}$ scheme:
\begin{eqnarray}
&&
^7{\rm F}_0 {\rm A}_{1g}^+ \, 0 {\rm cm}^{-1} \rightarrow 
^7{\rm F}_5 {\rm T}_{2g}^- \, 3153 {\rm cm}^{-1} \rightarrow 
^7{\rm F}_4 {\rm T}_{1g}^-\, 2266 {\rm cm}^{-1}
\,,
\end{eqnarray}
which requires two infrared(IR) lasers of frequencies, 390 meV and 109 meV,
or cascade type of excitation
\begin{eqnarray}
&&
^7{\rm F}_0 {\rm A}_{1g}^+ \, 0 {\rm cm}^{-1} \rightarrow 
^7{\rm F}_3 {\rm T}_{2g}^- \, 1554 {\rm cm}^{-1} \rightarrow 
^7{\rm F}_4 {\rm T}_{1g}^- \, 2266 {\rm cm}^{-1}
\,,
\end{eqnarray}
which requires two  lasers of frequencies, 192 meV and 88 meV.
Fabrication of IR lasers becomes more difficult when frequencies become smaller.
Another possibility is to irradiate two excitation lasers such that 
Raman type of excitation $\omega_1 - \omega_2 = \epsilon_{eg}$ occurs using lasers in
the optical region of $\omega_i\,, i = 1,2$.
A possible laser choice is $\omega_1 = 1064$ nm, $\omega_2 = 1402$nm.
On the other hand, Raman trigger at $|e \,, {\rm T}_{1g}^-\rangle \rightarrow | p \,, {\rm T}_{1g}^+\rangle $
is possible by single photon M1$^-$ transition, because
\begin{eqnarray}
&&
{\rm T}_{1g}^- \times \tilde{{\rm T}}_{1g}^- =  {\rm T}_{1g}^+ + {\rm T}_{2g}^+ + {\rm E}_g^+ + {\rm A}_{1g}^+ 
\,.
\end{eqnarray}

\begin{figure*}[htbp]
 \begin{center}
 \epsfxsize=0.6\textwidth
 \centerline{\includegraphics{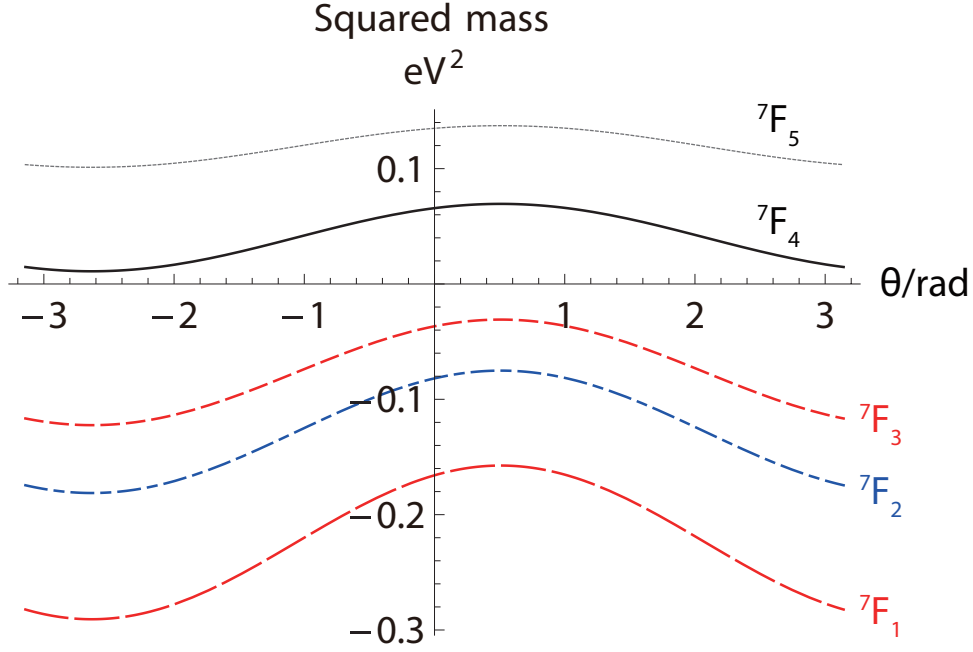}} \hspace *{\fill}
   \caption{
Squared mass function for $\theta_0 = 0.95 \pi$ in paths via possible J-manifold $|q\rangle $:
$^7$F$_4$ in solid black, $^7$F$_5$ in dotted black, $^7$F$_3$ in short dashed red,
$^7$F$_2$ in dash-dotted blue, and $^7$F$_1$ in long dashed magenda.
}
   \label{sm2+ mcq3rejection 2} 
 \end{center} 
\end{figure*}

A potential problem of rich 4f$^6$ level structure related to Sm$^{2+}$ scheme is 
whether the  McQ3 background  rejection is ensured in all  de-excitation paths.
In the squared mass function ${\cal M}^2(\theta)$ of eq.(\ref{squared mass f})
terms $\propto - \epsilon_{pq}$ become more negative and can become zero
at some scattered angle, when levels lower than adopted $| q\rangle $ are passed.
We check all squared mass function ${\cal M}^2(\theta)$ assuming
that states $| e\rangle \,, |p\rangle \,, |g\rangle$ are the same as
the specified path as above.
The result is shown in Fig(\ref{sm2+ mcq3rejection 2}), 
taking representative Stark levels in J-manifolds, 
$^7$F$_5$, $^7$F$_4$ (the given path),  $^7$F$_3$,
$^7$F$_2$, and $^7$F$_1$.
None of these vanish at any direction $\theta$: two of them are always positive
and three of them are always negative.
Negative squared mass functions imply absence of macro-coherent process,
while two positive ones contribute to RANP.
These two RANP processes are distinguishable by measuring Raman scattered
energy $\omega = \epsilon_{pq}$.

We have calculated squared mass functions for another excitation scheme using 
$|e \rangle $ belonging to $^7$F$_5$ manifold (allowed from the angular momentum point), which was found to have
lower de-excitation paths of contaminated McQ3 backgrounds.

Absolute RANP rates are numerically estimated using the formula,
\begin{eqnarray}
&&
\frac{d\Gamma_{RANP}} {d\Omega} = \frac{27 \pi}{8} 
\frac{\gamma_{pe}  \gamma_{pq} G_F^2}{\epsilon_{pe}^2\epsilon_{pq}^2 } 
\frac{n^4 V }{\Delta \omega_0 (\gamma_e^2 + \gamma_q^2 )^{3/2} }  
\eta | \tilde{\rho}_{qg}|^2 {\cal M}^2(\theta)
\,.
\end{eqnarray}
$  \tilde{\rho}_{qg}$ is coherence between the excited state $|q\rangle $ and $|g \rangle$.
At the moment it is not clear that a sizable $  \tilde{\rho}_{qg}$ is dynamically created by two
excitation lasers and Raman trigger laser, or that one needs extra laser or lasers for this purpose.
It seems necessary both to study this problem by dedicated simulations and to examine
it from experimental points.
Taking all decay rates, $\gamma_{pe},  \gamma_{pq},\gamma_e, \gamma_q $, to be 1 msec$^{-1}$,
$ n = 10^{16} {\rm cm}^{-3}\,, V = 10^{-2}  {\rm cm}^{3}\,, \Delta \omega_0 = 2\pi \times 1 {\rm GHz}$
gives
\begin{eqnarray}
&&
1.3 \times 10^7 {\rm sec}^{-1}\, 
\frac{{\cal M}^2(\theta) }{ \epsilon_{pe}^2\epsilon_{pq}^2} {\rm eV}^{2}\,
\eta | \tilde{\rho}_{qg}|^2 (\frac{n}{10^{16}{\rm cm}^{-3}})^4
\,.
\end{eqnarray}
Hence, the absolute rates are around $6 \times 10^9 $sec$^{-1}$  times 
$\eta  | \tilde{\rho}_{qg}|^2 {\cal M}^2(\theta)/{\rm eV}^2$.

We note that the original RENP ($| e \rangle \rightarrow | g\rangle + \gamma_0 + \nu \bar{\nu} $) rate 
stimulated by laser irradiation $\gamma_0$ is 
given by 
$3 G_F^2 \gamma_{pg} \epsilon_{eg} n^3 V \eta /(2 \epsilon_{pg}^3) $ times of 
order unity factors, \cite{renp rate}, \cite{renp overview},
which is completely negligible compared to RANP rate
and undetectable for the assumed excited target number density.

Finally, we shall estimate Rayleigh scattering rate.
Rayleigh scattering \cite{atomic physics text}, 
although not a background process against RANP due to different emitted photon
energy (Rayleigh scattering is low energy photon scattering against atoms/ions in the target without macro-coherence),
may cause serious damage to host crystals.
Taking SrF$_2$ as a host crystal, we estimate its host  density as $1.9 \times 10^{22}$cm$^{-3}$.
Using the refractive index of this crystal in the optical region, 1.4868, one may estimate Rayleigh scattering cross section
$\sim 1.4 \times 10^{-27}$cm$^2 \sin^2 \theta\, (\omega/{\rm eV} )^4$.
This gives, for 1mm$^2$ focused laser of 1mJ Rayleigh scattering rate, 
\begin{eqnarray}
&&
O(10^8) {\rm sec}^{-1} (\frac{\omega}{0.3 {\rm eV} })^4 \sin^2 \theta
\,.
\end{eqnarray}
This rate is roughly comparable to, or slightly less than, RANP rate.
The rate is  not very serious, but a care should be taken not to damage crystals.

\vspace{0.5cm}
\section
 {\bf Summary and prospects}

Symmetry is a compelling guiding principle of  challenging experiment of neutrino mass spectroscopy.
Assuming  as the target divalent lanthanoid ion at inversion center of O$_h$  host crystal symmetry,
we summarize results of Raman stimulated neutrino pair emission (RANP) as follows:

(1) Differential RANP rates are large; 
in the Sm$^{2+}:$SrF$_2$ example worked out in detail,
the rate is of order 
\begin{eqnarray*}
&&
d\Gamma =
6 \times (10^9 \sim 10^{10})  \,{\rm sec}^{-1} \eta  | \tilde{\rho}_{qg}|^2\,
\frac{{\cal M}^2(\theta)}{{\rm eV}^2}
\,(\frac{n}{10^{16}{\rm cm}^{-3}})^4 \frac{V}{10^{-2}{\rm cm}^3}\, d\Omega 
\,,
\end{eqnarray*}
taking a modest set of Raman trigger laser power and width.
Sm$^{2+}$ squared mass function $ {\cal M}^2(\theta)$ give
 rates of  order $  10^7  \,{\rm sec}^{-1} \eta   | \tilde{\rho}_{qg}|^2 $
per unit solid angle.
The rate can be raised by increasing Raman and coherence-generation
 laser power to obtain a larger excited target number density $n$ and the coherence factors.
There are considerable uncertainties of calculated RANP rates.
Most notably, the dominant M1 decay rates are not known with precision.
RANP rates depend on  product of decay rates $\gamma_{pe}\gamma_{pq} $, both of which were taken
1 msec$^{-1}$ in our estimate, but they should be determined by pilot experiments.

(2) Amplified QED backgrounds in Sm$^{2+}:$SrF$_2$ are less than
RANP rate, or even considering uncertainties of calculations, are made within controllable levels
by taking an appropriate range of Raman trigger angle in the specified scheme.

(3) Neutrino mass determination and Majorana/Dirac distinction become possible
by searching  threshold kinks in angular distribution of scattered Raman light.

Despite of these  merits RANP experiments might not be so easy.
Ideal perfect crystals never exist, and it is imperative to experimentally study
crystal qualities before we make a definite proposal of experiment.
The most important problem is whether divalent lanthanoid ions  in crystals
have an acceptable level of purity: whether defects  and impurities of host crystals
may not degrade the high symmetry required for RANP.
One of these check points may be to measure related background process
of phonon  induced extra photon emission in thermal media:
$\gamma_0 + \varphi_0 + |e \rangle \rightarrow \gamma + \gamma' + | g\rangle  $
with $\varphi_0$ a phonon in thermal equilibrium.
The process rate may be accelerated at will with increasing temperature.
By measuring rates at different temperatures one may be able to determine the background level.
Measurements of non-radiative relaxation should be studied by fabricated lanthanoid doped crystals.

\vspace{1cm}
 {\bf Acknowledgements}

We thank  F. Chiossi at Padova and Y. Kuramoto at KEK for valuable comments on lanthanoid doped crystals.
This research was partially
 supported by Grant-in-Aid 19K14741(HH), 19H00686(NS), and 17H02895(MY)  from the
 Ministry of Education, Culture, Sports, Science, and Technology.


\begin{thebibliography}{99}

\bibitem{pdg}
For a review of neutrino oscillation experiments,
Particle Data Group Collaboration, M. Tanabashi
et al., Phys. Rev. D 98 030001 (2018).

\bibitem{renp overview}
A. Fukumi et al.,
Prog.\ Theor.\ Exp.\ Phys.\ (2012) 04D002.

\bibitem{ranp}
H. Hara and M. Yoshimura,
arXiv: 1904.03813v1 (2019) and Eur.Phys.J. {\bf C79}  684(2019).

\bibitem{psr exp}
Y. Miyamoto et al., PTEP, 113C01 (2014).
Y. Miyamoto et al., PTEP, 081C01 (2015).
T. Hiraki et al., arXiv:1806.04005 [physics.atom-ph], and
J. Phys. B: At. Mol. Opt. Phys. {\bf 52}
045401 (2019).


\bibitem{psr th}
M. Yoshimura, N. Sasao, and M. Tanaka, Phys.
Rev.{\bf A86} 013812 (2012).


\bibitem{atomic physics text}
For a comprehensive textbook on atoms and molecules,
B.H. Bransden and C.J. Joachain, 
{\it Physics of Atoms and Molecules},
2nd edition, Prentice Hall(2003).



\bibitem{my-07}
M. Yoshimura, Phys. Rev. {\bf D75}, 113007(2007).


\bibitem{relic pauli blocking}
M. Yoshimura, N. Sasao, and M. Tanaka,
Phys.Rev.{\bf D 91}, 063516 (2015)


\bibitem{mcqn}
M. Yoshimura, N. Sasao, and M. Tanaka,
Prog. Theor. Exp. Phys. {\bf 2015}, 053B06 (1015).




\bibitem{van vleck}
J.H. Van Vleck, J. Chem. Phys. {\bf 41}, 67 (1937).




\bibitem{inui et al textbook}
T. Inui, Y. Tanabe, and Y. Onodera,
{\it Group Theory and its Applications in Physics},
Springer-Verlag, (1990).


\bibitem{rubio}
For a review of divalent lanthanoid ions doped in crystals, see
J. Rubio O.
J. Phys. Chem. Solids, {\bf 52}, 101 (1991).


\bibitem{judd-ofelt}
B.R. Judd, Phys.Rev.{\bf 127},  750(1962).
G.S. Ofelt, J. Chem. Phys. {\bf 37}, 511 (1961).






\bibitem{stark split oh}
K.R. Lea, M.J.M. Leask, and W.P. Wolf, J. Phys. Chem. Solids {\bf 23}, 1381 (1962).

\bibitem{ho2+ forced e1}
H.A. Weakliem and Z.J. Kiss, Phys.Rev. {\bf 157}, 277 (1967).








\bibitem{violated angular momentum}
As is well known and confirmed experimentally by violation of the optical selection rule in the free space, 
the rotational symmetry is broken down to point group symmetry
in crystals.
Nevertheless, in 4f systems the J-manifold concept is approximately
very useful since 4f electrons suffer least from crystal environment effects.


\bibitem{trivalent ion location}
For Er$^{3+}:$ LiYF$_4$, see
M.A. Couto dos Santos et al,,
J. of Alloys and Compounds, {\bf 275 - 277}, 435 (1998).

\bibitem{trivalent ion doped crystals}
A comprehensive review of trivalent lanthanoid ions doped in crystals is given 
from the lasing solid point of view by
M. Eichhorn, Appl. Phys. B {\bf 93}, 269-316 (2008).



\bibitem{sm2+  data}
 D.L. Wood  and W. Kaiser, Phys. Rev.{\bf 126}, 2079 (1962).

\bibitem{nu parameters}
Parameters determined from neutrino oscillation experiments \cite{pdg} and used in this work
are squared mixing matrix elements, $ |U_{ei}|^2, i = 1,2,3$, and
mass differences, $\delta m^2_{ij} = m_i^2 - m_j^2 $.
The smallest neutrino mass is assumed in each calculation.
CP violation phase factors, $\delta, \alpha, \beta$, are not known, and assumed
vanishing for simplicity, in the present work.

\bibitem{renp rate}
D. N. Dinh, S. Petcov, N. Sasao, M. Tanaka, and M. Yoshimura, Phys. Lett. {\bf B179}, 154 (2012).


\bibitem{srf2 debye}
L.T. Ho, D.P. Dandekar, and J.C. Ho,
Phys. Rev. {\bf B27}, 3881 (1983). 





%\bibitem{}

\end{thebibliography}
\end{document}